\documentclass[twocolumn]{aastex631}
\usepackage{ulem}
\usepackage{xcolor}
\usepackage[version=4]{mhchem}
\usepackage{savesym}
\savesymbol{tablenum}
\usepackage{siunitx}
\usepackage{booktabs}
\restoresymbol{SIX}{tablenum}
\usepackage{multirow}

\shorttitle{Oxirane + CH Radical reaction}
\shortauthors{Bensberg et al.}

\begin{document}

\title{Automated Exploration of Radical-Molecule Chemistry: The Case of Oxirane + \ce{CH} in the ISM}

\correspondingauthor{Silvia Alessandrini}
\email{silvia.alessandrini7@unibo.it}

\correspondingauthor{Cristina Puzzarini}
\email{cristina.puzzarini@unibo.it}

\correspondingauthor{Markus Reiher}
\email{mreiher@ethz.ch}

\author[0000-0002-3479-4772]{Moritz Bensberg}
\affiliation{ETH Zurich, Department of Chemistry and Applied Biosciences \& Center for the Origin and Prevalence of Life, Vladimir-Prelog-Weg 2, 8093 Zurich, Switzerland}
\author[0000-0003-3152-3261]{Silvia Alessandrini}
\affiliation{Dipartimento di Chimica ``Giacomo Ciamician'', Universit\`{a} di Bologna, Via P. Gobetti 85, Bologna, 40129, Italy}
\author[0000-0002-6492-5921]{Mattia Melosso}
\affiliation{Dipartimento di Chimica ``Giacomo Ciamician'', Universit\`{a} di Bologna, Via P. Gobetti 85, Bologna, 40129, Italy}
\author[0000-0002-2395-8532]{Cristina Puzzarini}
\affiliation{Dipartimento di Chimica ``Giacomo Ciamician'', Universit\`{a} di Bologna, Via P. Gobetti 85, Bologna, 40129, Italy}
\author[0000-0002-9508-1565]{Markus Reiher}
\affiliation{ETH Zurich, Department of Chemistry and Applied Biosciences \& Center for the Origin and Prevalence of Life, Vladimir-Prelog-Weg 2, 8093 Zurich, Switzerland}

\begin{abstract}
Quantum chemistry provides accurate and reliable methods to investigate reaction pathways of reactive molecular systems relevant to the interstellar medium. However, the exhaustive exploration of a reactive network is often a daunting task, resulting in unexplored reactive channels that affect kinetic outcomes and branching ratios. Here, an automated workflow for exploring reactive potential energy surfaces (PESs) is employed for the first time to study the oxirane (\ce{C2H4O}) plus methylidyne (\ce{^.CH}) reaction. The ultimate goal is to comprehensively map its PES and, subsequently, derive rate constants for the most important reaction channels. In addition to its astrochemical relevance, this reaction has been considered because it is a challenging test case, its network being very extended, with 60 exothermic bimolecular products lying below the reactant's energy. Kinetic simulations indicate that the main product of the reaction is the \ce{HCO} radical plus ethene (\ce{C2H4}), while formation of s-trans-propenal (acrolein) and 2H-oxene is also possible, but to a lesser extent. Based on the present study and other references in the literature, we suggest that the slightly higher relative abundance of s-trans-propenal compared to methyl ketene in the interstellar medium is a gas-phase kinetic effect, s-trans-propenal being a more easily accessible product on the \ce{C3H5O^.} PES. 

\end{abstract}


%%%%%%%%%%%%%%%%%%%%%%%%%%%%%%%%%%%%%%%%%%%%%%%%%%%%%%%%%%%%%%%%%%%%%%%%%%%%%%%%%%%%%%%%%%%%%%%%%%%%%%%%
\section{Introduction}\label{sec:introduction}
%%%%%%%%%%%%%%%%%%%%%%%%%%%%%%%%%%%%%%%%%%%%%%%%%%%%%%%%%%%%%%%%%%%%%%%%%%%%%%%%%%%%%%%%%%%%%%%%%%%%%%%%

The first discovery of interstellar molecules dates back to 1940, when McKellar identified the methylidyne radical (CH) and the cyano radical (CN) \citep{mckellar1940evidence}. To date, about 340 molecules have been detected, the vast majority being observed in the gas phase by radioastronomy. The census also includes organic molecules with a prebiotic character, and building blocks and cyano derivatives of polycyclic aromatic hydrocarbons (PAHs). The chemical complexity is undisputed as testified by the recent observation of isomers of cyanopyrenes \citep{wenzel2024detection,wenzel2025detections} and discovery of cyanocoronene \citep{Wenzel_2025}. In this respect, the CH and CN radicals, together with other small molecules, such as \ce{H2O}, \ce{H3+}, \ce{N2}, \ce{CO}, \ce{^.OH}, and \ce{^.CCH}, are considered building blocks of the chemical inventory that exists despite the harsh conditions of the interstellar medium, ISM (number density = $10-10^7~\text{cm}^{-3}$, gas temperature = $10-150~\text{K}$, ionizing radiation; \citet{yamamoto2017introduction,tielens2021molecular}). Hence, one of the main goals of astrochemistry is understanding the chemical reactivity of interstellar species, and how these have led to the observed chemical complexity  \citep{puzzarini2020grand}. In this regard, one can recall that reactivity occurs both on interstellar ices and in the gas phase. The former play an important role in the formation of organic molecules owing to the stabilization and catalytic effects offered by the ice surface. However, molecules to be detected by radioastronomy have to be desorbed and pass in the gas phase. This transition requires energy that is expected to be available only in warm and/or shocked regions \citep{jimenez2008parametrization,zeng2020cloud}. Differently, gas-phase neutral-neutral or ion-neutral bimolecular reactions are expected to be the dominant ones in the cold regions of ISM. 
Focusing on neutral-neutral reactions, they usually lead to bimolecular products, as no third-body stabilization of reaction intermediates occur in the rarefied interstellar gas \citep{puzzarini2022gas}. As little to no energy is available for these reactions at the low temperatures of the ISM, only exothermic processes are expected to take place, usually involving barrierless approaches of reactants and submerged barriers. However, tunneling might play a relevant role in the case of an emerged barrier (as in the case of the \ce{CH3OH + OH} reaction; \citet{ocana2019gas}).

Quantum chemistry offers the opportunity to investigate gas-phase reactions at a molecular level with high accuracy, thus overcoming the difficulties that experiments have in reproducing interstellar conditions. Quantum-chemical protocols have been developed to accurately investigate all reactive paths relevant to interstellar chemistry (e.g., limited to gas phase, \citet{jacs.7b12773,D0CP03953E,lupi2020state,alessandrini2021fate,recio2022intersystem,ctx5830378800007041,sciadv.adv0692,h2cs-h2co_cn}). However, in order to be informative, computational studies should be able to exhaustively explore all possible reaction mechanisms. This might require a significant (too large) human effort. To solve this issue, in recent years, several computational tools for the automated exploration of reactive potential energy surfaces (PESs) have been introduced (see, e.g., 
\citet{Maeda2005,Zimmerman,Bergeler2015,martinez2015automated1,martinez2015automated2,jctc.9b01006,Simm2017,jctc.3c00752,jpca.3c05253}). However, their application to interstellar reactions has been found unsatisfactory because of reactive channels that remain unexplored, thus affecting kinetic outcomes and branching ratios \citep{tonolo2020quest,west2023experimental,h2cs-h2co_cn,puzzarini2020twist}. At the same time, the pace of molecules detection in the ISM has grown rapidly \citep{McGuire_2022}, with experimental and theoretical studies of formation reactions and their rates that struggle to keep up, resulting in astrochemical models that rely on assumed rate constants \citep{wakelam20152014,tinacci2023gretobape,millar2024umist}. Therefore, a full automation of the process for deriving a reactive PES, followed by the calculation of kinetic rate coefficients, is essential to determine reliable data and update models at a rapid pace. To address this challenge, this work aims to present a fully automated workflow for exploring the PES of astrochemically relevant reactions, with that occurring between the CH radical and oxirane (ethylene oxide, \ce{c-C2H4O}) being selected as test case in view of its complexity. Both reactants have been identified in the ISM. The CH radical is indeed ubiquitous \citep{awad2022interstellar,yamamoto2017introduction}, while oxirane has been observed in the Center Molecular Zone \citep{dickens1997detection,requena2008galactic}, star-forming regions \citep{nummelin1998abundances}, and the protostar IRAS 16293 \citep{lykke2017alma}. Other than these hot sources, oxirane has recently been detected in colder regions like TMC-1 \citep{barnum2022search} and the prestellar core L1689B \citep{bacmann2019cold}. Thus, the investigated reaction is relevant across different astronomical sources and stages of star formation. 

As mentioned above, the title reaction was selected as it represents a challenging test case for automated network exploration for multiple reasons. First, \ce{^.CH} ($X^2\Pi$) is a highly reactive radical, with an unpaired electron, a vacant non-bonding molecular orbital on the carbon atom and a pair of electrons in the 2$s$ state that can easily provide access to a quartet state.  This peculiarity allows both addition and abstraction reactions, even under interstellar conditions \citep{doddipatla2021low,krikunova2023reaction,caster2019kinetic,caster2021product,he2022chemical}. Second, oxirane represents an unstable species belonging to the \ce{C2H4O} isomeric family. Having a three-membered ring structure \citep{karton2014pinning}, it offers three different approaches to the CH radical: (i) the oxygen atom, (ii) the two equivalent carbons, and (iii) the four equivalent hydrogen atoms. Hence, a large variety of entrance channels is possible, with several products and transitions states expected to lie below the reactants' asymptote due to the instability of the reactive pair. In detail, the \ce{C3H4O} isomers are expected to be relevant products on the CH + \ce{c-C2H4O} reactive PES. These would correspond to the addition of CH radical followed by a H atom loss, with species like propenal, methyl ketene, and cyclopropanone being potential products. Surprisingly, the most stable species of this isomeric family has long been disputed, with methyl ketene resulting to be the real global minimum \citep{field2021conjugation}. Therefore, according to the minimum energy principle (MEP; \citet{lattelais2009interstellar}), this species is suggested to be the most abundant in the ISM. However, recent findings on the abundance of s-trans-propenal and methyl ketene in TMC-1 indicate that the former is more abundant than the latter \citep{fuentetaja2023discovery}. This might be related to a kinetic effect which has been explained by exploiting the rates for the \ce{^.OH} + \ce{CH3CCH} \citep{taylor2008pulsed} and \ce{CH3CHO + ^.CH} reactions \citep{goulay2012product,wang2017theoretical}. Since oxirane has recently been detected in TMC-1 \citep{barnum2022search}, the title reaction can help understand the observed relative abundances of these two species. Focusing on other possible products, in addition to \ce{H^.} and \ce{H2}, \ce{^.CH3} and \ce{HCO^.} are expected to be small co-products of the title reaction. The elimination of the \ce{CH3} radical would result in the formation of \ce{C2H2O} isomers, suggesting a dehydrogenation process caused by the \ce{CH} radical. Instead, the loss of \ce{HCO^.} would represent a destruction pathway, with the formation of ethene (\ce{C2H4}). The present work offers the opportunity to assess these different possibilities and compare the results with those issuing from the \ce{CH3CHO + ^.CH} reaction, where acetaldehyde is a structural isomer of oxirane, and from the \ce{CH + c-C5H6} reaction, which is representative of ring expansion due to \ce{^.CH} addition.  

Thi manuscript is structured into three main sections. First, the computational methodology is explained with a focus on the strategy to exhaustively explore reactive PESs. Then, the key thermodynamic and kinetic results are presented and discussed in Section 3. Finally, a section summarizing the main conclusions is provided. 

%%%%%%%%%%%%%%%%%%%%%%%%%%%%%%%%%%%%%%%%%%%%%%%%%%%%%%%%%%%%%%%%%%%%%%%%%%%%%%%%%%%%%%%%%%%%%%%%%%%%%%%%
\section{Methodology}\label{sec:methods}
%%%%%%%%%%%%%%%%%%%%%%%%%%%%%%%%%%%%%%%%%%%%%%%%%%%%%%%%%%%%%%%%%%%%%%%%%%%%%%%%%%%%%%%%%%%%%%%%%%%%%%%%

Exploring the gas-phase reaction between oxirane and the CH radical using quantum chemical approaches implies the identification of all relevant (low-energy) transition states, intermediates, and potential products. Traditionally, the identification of these stationary points is a labor-intensive task which is performed manually by a quantum chemist. However, the increase in computational resources and progress in software development has made it possible to automate this task. Currently, various automated reaction network exploration approaches are available (see, for example, \citet{Dewyer2017,Simm2018,Green2019,Maeda2021,Steiner2022} for reviews and references cited therein).

One key challenge in automated reaction network exploration is the description of the reaction mechanism, which is often based on identifying transition states on the PES.
(variant NT2;~\citet{Unsleber2022,Quapp2020}) available in the SCINE framework~\citep{Weymuth2024a}. The NT2 algorithm is used in the Chemoton exploration software~\citep{Unsleber2022, Bensberg2024e} to scan a PES for guesses of transition state structures along a potential reaction coordinate defined through reactive atom pairs. A key feature of this approach is its single-ended strategy, not requiring any information about possible reaction products. However, if a guess for the products is available, double-ended approaches can be used to identify transition states connecting products and reactants by, for instance, the double-ended variant of the growing string method~\citep{Zimmerman2013a} or through curve optimization~\citep{Vaucher2018}.
The products required for these interpolation techniques can be obtained from molecular dynamics simulations as done in the Nanoreactor approach~\citep{Wang2014}, using hyperdynamics~\citep{StanBernhardt2024}, reactive molecular dynamics~\citep{Ensing2006,Martinez‐Nunez2021,zhang2025exploring}, 
or other systematic explorations as done in CREST~\citep{Pracht2024} for nonreactive PESs 
and by the QCxMS2 method for reactive networks~\citep{Gorges2025a}. Graph enumeration schemes are often employed as a low-cost alternative~\citep{Zhao2021a}.

In this work, exploiting the NT2 algorithm in Chemoton~\citep{Unsleber2022, Bensberg2024e} we explore the PES of the reaction between oxirane and the CH radical. We rely on complete automation of the exploration procedure in Chemoton, requiring only the reactant structures as input. We use distributed computing to explore multiple reaction paths simultaneously~\citep{Weymuth2024a}. Chemoton utilizes the program Molassembler~\citep{Sobez2020, Bensberg2024f} to characterize any newly discovered minimum structure on the PES using its molecular graph and local geometry at the individual atoms, thereby differentiating enantiomers. These minimum \textit{structures} are grouped into \textit{compounds}, according to their Molassembler characterization, charge, and multiplicity. Therefore, 'compounds' provide a systematic bookkeeping of the discovered species. The reaction discovery calculations in Chemoton not only search for transition states, but also optimize intrinsic reaction coordinates (IRCs;~\citet{fukui1981path}) to identify a minimum energy path connecting reactants and products. Based on these minimum energy paths, we define \textit{elementary steps} that connect reactant and product structures. Based on these elementary steps, we define \textit{reactions} as the transformation between compounds, as described by the elementary steps that transform the structures associated with the compounds \citep{Unsleber2020}.

We extend the PES exploration to model reactivity under the conditions typical of the ISM: very low pressure and very low temperature. Therefore, we chose the following conditions for deciding which intermediates to explore further:

(i) Energy limitation.
The total ZPE-corrected energy $E_i$ of the species $i$ (ZPE standing for zero-point energy) must be lower than the energy $E_\mathrm{reactant}$ of the initial reactants (oxirane and the CH radical). Furthermore, there must be a reaction path $P$ from the reactants to $i$ that does not cross a transition state or intermediate $j$ with an energy $E_j$ higher than $E_\mathrm{reactant}$:
\begin{align}
    E_i < E_\mathrm{reactant} ~~\mathrm{and}~~ \max_P\left(\max_{j\in P}(E_j)\right) < E_\mathrm{reactant}
    \label{eq:energy_cut_off}
\end{align}
This constraint is imposed by the low temperature of the ISM.

(ii) Bimolecular reaction restriction.
Reaction products should be bimolecular, which means that the reaction path proceeds until formation of a bimolecular product. The reason for this constraint is that the excess of energy cannot be removed by third-body effects (due to the very low pressure of the ISM).

(iii) Reaction coordinate selection.
All possible reaction coordinates are explored, with changes of at most two bonds at a time (one bond formation and one bond dissociation) being considered.

These simple criteria allow for an exhaustive PES exploration, which is typically impossible in solution or high-pressure chemistry, where kinetic guidance is required~\citep{Bensberg2023a, Liu2021, Woulfe2025}. Nevertheless, a large number of exploratory reaction-discovery calculations was needed. Therefore, we reduced the computational cost by employing a refinement-based approach. The molecular structures were calculated using the computationally efficient  PBE~\citep{Perdew96} exchange-correlation energy density functional, incorporating the D3 dispersion correction~\citep{Grimme2010a} and Becke--Johnson damping~\citep{Grimme2011} (denoted as PBE-D3 in the following), in conjunction with the def2-SVP basis set~\citep{Ahlrich2005}. Afterwards, key electronic energies were refined by explicitly correlated, domain-based local pair natural orbital coupled cluster single-point calculations with single, double, and perturbative triple excitations, DLPNO-CCSD(T)-F12~\citep{Kumar2020}, in the cc-pVDZ-F12 basis set~\citep{Peterson2008}. This energy refinement was applied only to those transition states or minimum structures whose PBE-D3/def2-SVP ZPE-corrected energy was lying below that of reactants (oxirane + \ce{^.CH}), with a tolerance of 40~kJ/mol. Furthermore, only the 10 most favorable transition states of a reaction were refined if lying within an energy window of 20~kJ/mol with respect to the most stable transition state. The same protocol was applied to minimum structures. Note that we always used the DLPNO-CCSD(T)-F12/cc-pVDZ-F12 energy, including the PBDE-D3/def2-SVP zero point correction to evaluate the energy limitation criterion expressed by Eq.~\ref{eq:energy_cut_off}.

The Master Equation System Solver (MESS) program~\citep{Georgievskii2013} was used to determine the kinetic rate coefficients for the bimolecular channels observed in the network. For the collisional model, Ar gas was considered as the collider and the collisional parameters were derived from \citet{jasper2020third}, considering oxirane similar to alcohols. The kinetic simulation was run in the 40-500 K temperature range, in step of 5~K in the 40-200~K interval and of 50~K between 200 and 500~K. The pressure dependence of the reaction was also checked in the $1 - 1\times 10^{-7}$~atm range. The input for the kinetic simulation is available on Zenodo \citep{zenodoArchive}. 
Aiming at bimolecular channels, reactions leading to further fragmentation of the bimolecular products, discovered during the exploration, were excluded from the kinetic modeling. As mentioned above, only the reactions that did not lead to bimolecular products or did cross transition states requiring a total energy higher than the reactant energy $E_\mathrm{reactant}$ were excluded from the master equation (ME) 
simulation. The first step of the reaction, that is the barrierless entrance channel, was modeled using the phase space theory (PST; \citet{chesnavich1986multiple,pechukas1965detailed}). The spin-orbit coupling of between the $X_{1}~^2\Pi_{1/2}$ and $X_{2}~^2\Pi_{3/2}$ states of the \ce{CH} radical was also considered in the input of the ME simulation, using 26~cm$^{-1}$ as splitting value according the literature \citep{song2008potential}. For each relevant product, the temperature dependence of the rate constants has been modeled using the Arrhenius-Kooij formula \citep{kooij}:
\begin{equation}
k(T)=\alpha \left( \frac{T}{300} \right) ^{\beta} \text{exp}{\left( -\frac{\gamma}{T} \right)}
\label{eq72}
\end{equation}
where the fitting parameters are three: $\alpha$ which represents a pre-exponential factor (cm$^3$ molecule$^{-1}$ s$^{-1}$), $\beta$ (dimensionless) that accounts for the temperature dependence of the pre-exponential factor, and $\gamma$ (K) which is the effective activation energy. 

%%%%%%%%%%%%%%%%%%%%%%%%%%%%%%%%%%%%%%%%%%%%%%%%%%%%%%%%%%%%%%%%%%%%%%%%%%%%%%%%%%%%%%%%%%%%%%%%%%%%%%%%
\section{Results and Discussion}\label{sec:results}
%%%%%%%%%%%%%%%%%%%%%%%%%%%%%%%%%%%%%%%%%%%%%%%%%%%%%%%%%%%%%%%%%%%%%%%%%%%%%%%%%%%%%%%%%%%%%%%%%%%%%%%%

\subsection{Reaction Network Overview and Entrance Channels}

The reaction network exploration terminated naturally after a few days, using 900 cores in parallel, as soon as no new compounds fulfilling the exploration criteria introduced above were discovered. An overview of the final size of the reaction network is given in Table~\ref{tab:network_overview}. In total, we required 17927 NT2-type reaction exploration calculations to sample the PES, resulting in 9313 elementary steps, 818 reactions, and 212 compounds. The energy restriction for the DLPNO-CCSD(T)-F12/cc-pVDZ-F12 refinement proved very effective in reducing the total number of single-point calculations. Indeed, starting from a total of 37385 transition states and minimum structures, selection criteria reduced the energy refinement to only 5329 single points.

\begin{table}[t]
    \centering
    \caption{Number of entries for key quantities in the final database representing the reaction network.}    \label{tab:network_overview}
    \begin{tabular}{l r}
    \toprule\toprule
        NT2 Calculations & 17927 \\
        CC Single Points & 5329 \\\midrule
        Compounds & 212 (175)\textsuperscript{a} \\
        Compound Complexes & 33 (29)\textsuperscript{a} \\
        Reactions & 818 \\
        Elementary Steps & 9313 \\
        Structures\textsuperscript{b}& 37385\\
    \bottomrule\bottomrule
    \end{tabular}
    \footnotesize
    \\
    \textsuperscript{a}{Number of compounds/compound complexes fulfilling the energy criterion in Eq.~\ref{eq:energy_cut_off} }
    \textsuperscript{b}{Structures refer to minima and transition states on the PES. The total number includes duplicates generated independently by separate NT2 calculations.}
\end{table}

\begin{figure}
    \centering
    \includegraphics[width=0.8\linewidth]{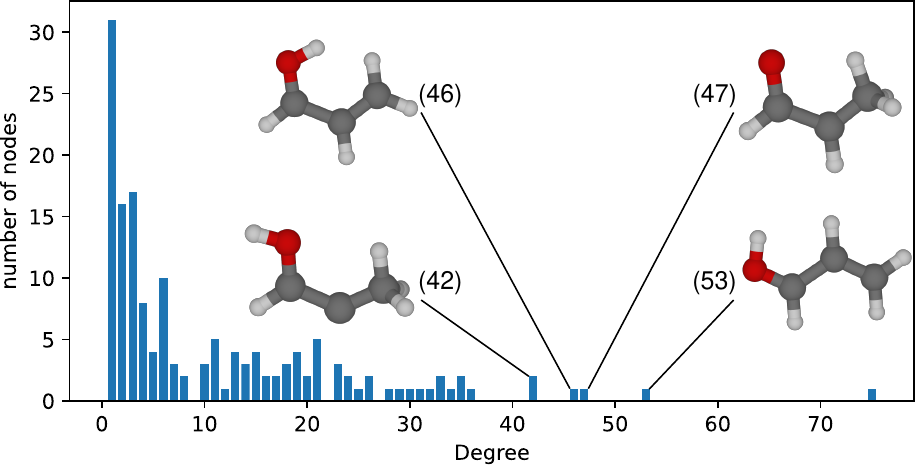}
    \caption{Degree distribution, i.e., number of incoming + outgoing reactions for each compound, of the reaction network. The molecules corresponding to the highest degrees, namely 42, 46, 47, and 53, are depicted. The  degree of 75 corresponds to the hydrogen atom. The second molecule with degree 42 is \ce{H2}.}
    \label{fig:degree_distribution}
\end{figure}

The large number of reactions (818) compared to the number of compounds (212) suggests that the core of the reaction network is highly connected. The degree distribution of the reaction network graph, which is the number of incoming and outgoing reactions for each species and shown in Fig.~\ref{fig:degree_distribution}, indicates the high connectivity of a small number of core species. However, most of the compounds correspond to fragmentation reaction products, which were not explored further, thus resulting in a small degree. Notably, the hydrogen atom has a very high degree, 75, because it is part of a large number of fragmentation reactions. 

\begin{figure*}
    \centering
    \includegraphics[width=1\linewidth]{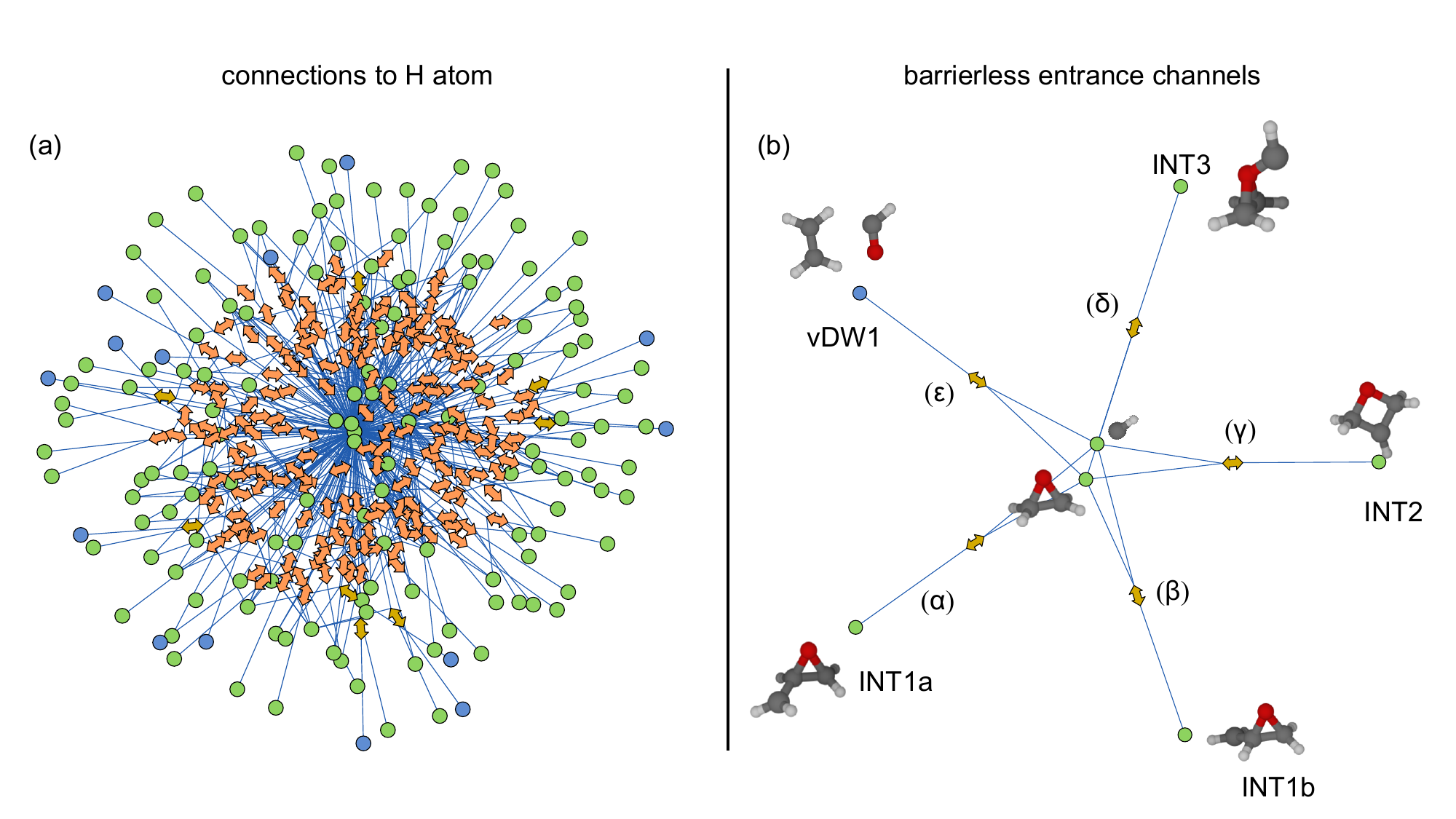}
    \caption{Panel (a) reports the congested network of the reaction between oxirane and the \ce{CH} radical considering at the center of the figure the most connected species, the H atom. Green points indicate minima (exit products or intermediates), while blue points are weakly bound complexes. The orange arrows indicate the IRC connecting the different intermediates. Panel (b) shows the barrierless approaches between the two reactants and the intermediates/complexes formed.}
    \label{fig:global_entrance}
\end{figure*}

The vast dimension of the PES does not allow for an easy representation, so the reader is referred to the Zenodo archive \citep{zenodoArchive} for its visualization as well as to panel (a) of Fig.~\ref{fig:global_entrance}, where the reactions involving the most connected species (the H atom) are shown. This represents the opening interface of SCINE Heron \citep{Mueller2024a, Bensberg2024g}, which can be used to navigate the network. Overall, the kinetic modeling with MESS considered 60 exothermic bimolecular products and about the same amount of intermediates, which are shown in Appendices \ref{app_bimolecular} and \ref{app_compounds}, respectively. The loss of a H atom results in 26 of the 60 possible products, thus having the formation of 26 different \ce{C3H4O} isomers. The loss of an \ce{H2} molecule leads to 23 different products (23 isomers of the \ce{C3H3O} radical). The elimination of the \ce{CH3} radical results in four different products, while \ce{CH4} is the co-fragment of 2 products. The \ce{HCO} and \ce{HOC} radicals are co-fragments of ethene, thus accounting for two products. The \ce{H2CO} + \ce{^.C2H3} product is also observed on the PES. The last two products are \ce{H2O + CH2CCH^.} and \ce{CH3CH2^. + CO}. The latter is also the most exothermic product (P38), being located at -536 kJ/mol with respect to the reactants. The \ce{HCO^. + C2H4} pair is the second most exothermic product, located at -445 kJ/mol. Very close in energy are P49 (-442 kJ/mol, \ce{^.CH3 + H2CCO}), P20 (-440 kJ/mol, \ce{CH2CHCO^. + H2}), and P34 (-429 kJ/mol, \ce{H2O + CH2CCH^.}).

As expected, the entrance channels of the reaction are several, but we did not observe a direct H-abstraction resulting from the CH radical. Panel (b) of Fig.~\ref{fig:global_entrance} shows the barrierless connection from reactants to the intermediates. Approaches ($\alpha$) and ($\beta$) show the attack of \ce{^.CH} on one of the \ce{CH2} vertices of oxirane. In this case, a H atom is abstracted to form the \ce{CH2} radical, which further reacts in a concerted manner. This leads to INT1a and INT1b, which represent the simple substitution of a H atom with a \ce{CH2} group. As shown in panel (b) of Fig.~\ref{fig:global_entrance}, INT1a and INT1b simply show different orientations of the \ce{CH2} terminal group in space. Approach ($\gamma$) is the addition of the \ce{CH} radical to the C-C bond of oxirane, resulting in the formation of a 4-membered cyclic intermediate (INT2). Lastly, the radical addition to the oxygen atom of oxirane is the approach ($\delta$) and leads to INT3, while a barrierless process directly forming a weakly bounded van der Waals complex (vdW1) between ethene and \ce{HCO^.}, denoted vDW1, is observed via approach ($\varepsilon$). The latter then evolves into P40, i.e. the \ce{HCO^. + C2H4} product. The scans employed to model these entrance channels using PST are reported in Appendix \ref{app_entrancechannels}.

\subsection{Paths evolution and Kinetics}

\begin{figure*}
   \centering
   \includegraphics[width=0.8\linewidth]{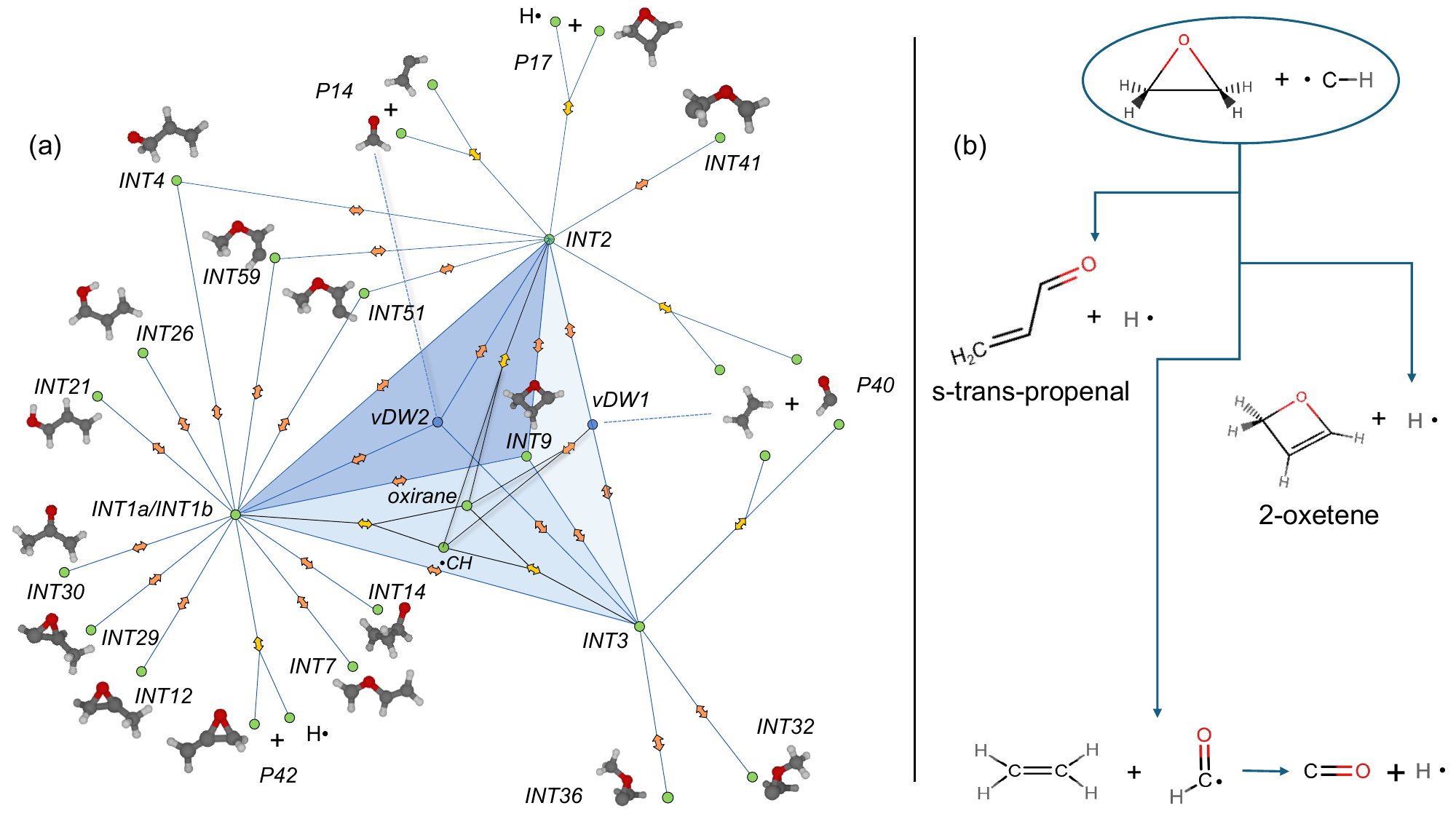}
   \caption{Panel (a) shows the interconnection among the four entrance intermediates (INT1a/INT1b, INT2, INT3) and the products directly accessible from them. The entrance channels (already shown in Fig.~\ref{fig:global_entrance}) are in black, while the new channels are in blue. The dashed lines indicate a barrierless dissociation from vdW wells to products. For simplicity, the species are indicated with the label introduced in Figs.~\ref{figA1}, \ref{figA2}, and \ref{figA3} of the Appendix. Panel (b) reports the three main products of the reaction according to the present kinetic simulation.}
    \label{fig:intermediates}
\end{figure*}

Figure \ref{fig:intermediates} (panel (a)) shows the different connections arising from the entrance intermediates. It is interesting to note that the species are all connected to each other by one-step processes, i.e. ruled by one single IRC, apart from the pair INT2/INT3. These can only interconvert through species 9. Both INT2 and INT3 can evolve into vDW1 (introduced above), which leads to P40 (\ce{HCO^. + C2H4}). This product can also be formed directly from both intermediates, without going through the formation of the weakly interacting complex. From INT2 other two pairs of products can be formed: \ce{H2CO + CH2CH^.} (P14) and 2H-oxetene (\ce{c-C3H4O}) plus a H atom (P17). The former product can be obtained from vDW2, accessible from INT2, INT3 and INT1a/INT1b. 

Apart from the one-step processes, INT2 paths are very tangled with those arising from INT1a/INT1b. Indeed, they share four intermediate species: INT4, INT59, INT51 and INT9. The only path that, in the picture, does not reconnect in one or two steps to INT1a/INT1b are: INT3 $\rightarrow$ species INT36, species INT32 and INT2 $\rightarrow$ species INT41.

From INT1a/INT1b, P42 (\ce{CH2=c-C2OH3 + H^.}) can be obtained. Furthermore, pathways toward the intermediates having the \ce{CH2=CH\bond{~-}CH\bond{~-}OH^.} formula (INT21 and INT26) are observed in Fig.~\ref{fig:intermediates}. These closely resemble trans-propenal (acrolein), and they are among the intermediates with the largest number of connections ($>40$) on the PES, as shown in Fig.~\ref{fig:degree_distribution}. These can be considered key intermediates of the reactions. The connections above this level become too tangled to be represented, indeed not even half of the minima of the reaction are shown in Fig.~\ref{fig:intermediates}. For this reason we aim to describe only the relevant pathways according to the kinetic simulations.

To establish the outcome of the oxirane + \ce{^.CH} reaction, kinetic simulations using MESS have been carried out in the 40-500~K range in the low-pressure limit ($1 \times 10^{-7}$~atm). The full list of the rate coefficients obtained at each temperature considered is reported in Table \ref{table:kinetics} for the main products of the title reaction, while in panel (b) of  Fig.~\ref{fig:intermediates} the most important products are sketched. The rates show no pressure dependence in the range 1-$1 \times 10^{-7}$~atm (see Appendix \ref{pressure}), while different temperature dependencies are observed for different products. The main outcome of the title reaction is a destruction of the oxirane ring, this leading to \ce{HCO^. + C2H4} (P40). This was somewhat expected as this product can be formed without barrier in the entrance channel. The second-fastest rate is that for the formation of s-trans-propenal + \ce{H^.} (P18), followed by 2H-oxetene + \ce{H^.} (P17). To give an idea of the temperature dependence, the formation rate of \ce{HCO^. + C2H4} shows a small decrease by increasing temperature due to vDW1 being formed before the bimolecular product. A similar trend is observed for s-trans-propenal while, for the formation rate of 2H-oxetene, a slightly increase when increasing the temperature is noted. The formation of 2H-oxetene becomes comparable to that of acrolein at T $>$ 300~K. The branching ratio for the formation of \ce{HCO^. + C2H4} is expected to be 85-86\% in the temperature range considered, while that for trans-propenal is around 8-9\%. 

As mentioned in the Methodology section, the rate coefficients obtained from the kinetic simulations have been fitted using a modified Arrhenius equation. The parameters are collected in Table \ref{table:kinetics}, while the fits are available in Appendix \ref{pressure} and can be used to extrapolate the rate constants at lower temperatures.

\begin{table*}
    \small
    \centering
    \caption{Temperature dependence of the rate coefficients for the relevant pathways of the oxirane + \ce{CH^.} reaction }\label{table:kinetics}
    \begin{tabular}{l c c c c }
\hline\hline
      temperature / K &  \multicolumn{4}{c}{$k$ / cm$^3$ molecule$^{-1}$ s$^{-1}$}                                    \\ \cline{2-5}
   &  Ethene + \ce{HCO^.} & 2H-oxetene + \ce{H^.} & s-trans-propanal + \ce{H^.} &  Methyl ketene + \ce{H^.} \\ \hline                                                                        
40      &      7.41E-10        &    3.57E-11           &    7.79E-11               &     4.83E-12              \\
45      &      7.22E-10        &    3.53E-11           &    7.59E-11               &     4.70E-12              \\
50      &      7.07E-10        &    3.50E-11           &    7.43E-11               &     4.60E-12              \\
55      &      6.96E-10        &    3.49E-11           &    7.30E-11               &     4.52E-12              \\
60      &      6.85E-10        &    3.47E-11           &    7.18E-11               &     4.44E-12              \\
65      &      6.76E-10        &    3.47E-11           &    7.08E-11               &     4.38E-12              \\
70      &      6.69E-10        &    3.46E-11           &    7.00E-11               &     4.33E-12              \\
75      &      6.62E-10        &    3.46E-11           &    6.93E-11               &     4.28E-12              \\
80      &      6.57E-10        &    3.46E-11           &    6.87E-11               &     4.24E-12              \\
85      &      6.52E-10        &    3.47E-11           &    6.82E-11               &     4.21E-12              \\
90      &      6.47E-10        &    3.47E-11           &    6.76E-11               &     4.17E-12              \\
95      &      6.44E-10        &    3.48E-11           &    6.73E-11               &     4.15E-12              \\
100     &      6.40E-10        &    3.48E-11           &    6.68E-11               &     4.12E-12              \\
105     &      6.38E-10        &    3.49E-11           &    6.65E-11               &     4.10E-12              \\
110     &      6.34E-10        &    3.49E-11           &    6.62E-11               &     4.08E-12              \\
120     &      6.30E-10        &    3.52E-11           &    6.57E-11               &     4.04E-12              \\
125     &      6.28E-10        &    3.52E-11           &    6.54E-11               &     4.02E-12              \\
130     &      6.26E-10        &    3.53E-11           &    6.51E-11               &     4.01E-12              \\
135     &      6.25E-10        &    3.55E-11           &    6.50E-11               &     4.00E-12              \\
140     &      6.23E-10        &    3.56E-11           &    6.48E-11               &     3.99E-12              \\
145     &      6.21E-10        &    3.57E-11           &    6.46E-11               &     3.97E-12              \\
150     &      6.20E-10        &    3.58E-11           &    6.44E-11               &     3.96E-12              \\
200     &      6.12E-10        &    3.70E-11           &    6.34E-11               &     3.88E-12              \\
250     &      6.07E-10        &    3.84E-11           &    6.26E-11               &     3.82E-12              \\
300     &      6.00E-10        &    3.97E-11           &    6.19E-11               &     3.76E-12              \\
400     &      5.92E-10        &    4.27E-11           &    6.06E-11               &     3.64E-12              \\
450     &      5.87E-10        &    4.43E-11           &    5.98E-11               &     3.57E-12              \\
500     &      5.80E-10        &    4.61E-11           &    5.90E-11               &     3.50E-12              \\
\hline
\multicolumn{5}{c}{Arrenihus-Kooji fit parameters}                                                                     \\
\hline
$\alpha$ (cm$^3$ molecule$^{-1}$ s$^{-1}$)   &   5.794E-10 $\pm$ 7.4E-13 &     3.799E-11 $\pm$ 1.7E-13  &   5.970E-11 $\pm$ 9.2E-14 &  3.63E-12 $\pm$ 8.7E-15 \\
$\beta$    &   -0.012 $\pm$ 0.003      &     0.258 $\pm$ 0.009        &   -0.030 $\pm$ 0.003      &  -0.058 $\pm$ 0.005     \\
$\gamma$ (K)  &   -8.86 $\pm$ 0.28        &     -19.07 $\pm$ 1.01        &   -8.16 $\pm$ 0.34         &  -6.55 $\pm$ 0.52       \\
rms        &  1.95E-12                 &         3.99E-13             &   2.46E-13                &  2.32E-14               \\
\hline\hline
    \end{tabular}
\end{table*}

\subsection{Discussion}

\begin{figure}
    \centering
    \includegraphics[width=0.98\linewidth]{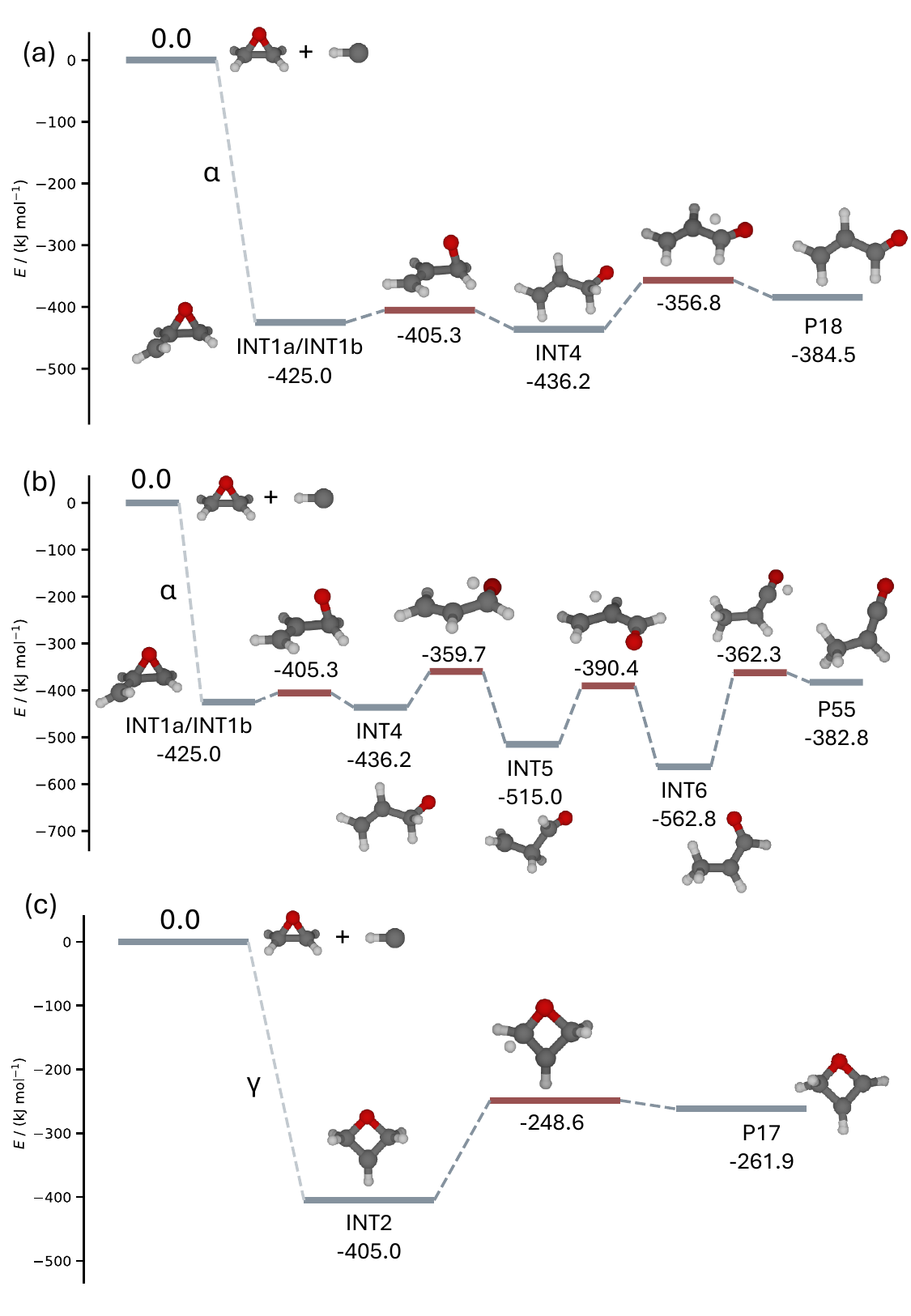}
    \caption{Shortest reaction path according to Pathfinder for the formation of (a) s-trans-propenal (P18), (b) methyl ketene (P55), and (c) 2H-oxetene (P17).}
    \label{fig:H-elimination-paths}
\end{figure}

Kinetic simulations indicate that the destruction path towards \ce{HCO^. + C2H4} is dominant, while the formation rate of s-trans-propenal is about one order of magnitude smaller, with a branching fraction of about 8-9\%. The reaction also leads to the formation of methyl ketene, whose rate coefficients are also reported in Table \ref{table:kinetics}. It is noted that they are about one-order of magnitude smaller than those of trans-propenal. For example, the rate constants are 4.12$\times 10^{-12}$~cm$^3$ molecule$^{-1}$ s$^{-1}$ for the pathway to methyl ketene and 6.68 $\times 10^{-11}$~cm$^3$ molecule$^{-1}$ s$^{-1}$ for the formation of s-trans-propenal at T = 100 K. To understand the differences in the formation of these two species, the shortest reaction paths according to Pathfinder \citep{Tuertscher2022a} are plotted in Fig.~\ref{fig:H-elimination-paths}: panel (a) for trans-propenal and panel (b) for methyl ketene. From their comparison, it is evident that trans-propenal and methyl ketene share the first step towards their formation: the system evolves in one step from INT1a/b to INT4 with a barrier of 20 kJ/mol. From this intermediate, trans-propenal is obtained in a single step by overcoming a barrier of about 80 kJ/mol. Instead, the formation of methyl ketene requires additional steps. From INT4, three different barriers need to be overcome. Among these, the first is still comparable to that for the formation of trans-propenal ($\sim$ 80 kJ/mol), but the other two barriers are by far higher, 125 and 200 kJ/mol, thus significantly reducing the reaction rate. The fastest formation of trans-propenal can also be explained considering the fact that the most connected species in the network resemble trans-propenal and could lead to its formation by simple cleavage of a C-H bond and subsequent elimination of a H atom. 

In panel (c) of Fig.~\ref{fig:H-elimination-paths}, the path towards 2H-oxetene is shown. This species has a production rate comparable to that of trans-propenal, with the shortest pathway starting from the formation of INT2 followed by H atom elimination. 2H-Oxetene is a cyclic structure that is known to photo-isomerize to form trans-propenal under UV irradiation in the 4.9-5.2~eV range \citep{Kikuchi1981}. Being this cyclic species very unstable, its isomerization can easily occur in presence of a catalytic surface or any energetic source. Thus, one might expect that, in the ISM, if 2H-oxetene is formed, it eventually transforms into trans-propenal. Interestingly, no rotational spectroscopic data are available for this molecule, thus one can only speculate about its presence until the experimental data required for astronomical searches will be available. 

Overall, the title reaction compares well with the \ce{CH3CHO + ^.CH} reactive system (\ce{CH3CHO} being an isomer of oxirane), whose PES exploration however has only been partially investigated. Therefore, no conclusions can be drawn on reaction rates and branching ratios \citep{wang2017theoretical}. Experimentally, several products have been observed \citep{goulay2012product}. Among them, the main product belonging to the \ce{C3H4O} family is propenal, followed by methylene-oxirane and methyl ketene. However, in the oxirane plus \ce{^.CH} reaction, the methylene-oxirane + H (P42) formation is slower than that of methyl ketene, the corresponding rate constant being on the order of $3\times 10^{-13}$ cm$^3$ molecule$^{-1}$ s$^{-1}$ in the temperature range considered. Since trans-propenal is a more abundant product than methyl ketene in both reactions, this tends to suggest that the slight difference in the abundance ratio observed in the ISM arises from a kinetic effect related to the \ce{C3H5O} reactive PES. Another interesting outcome of this study is that ring-extension mechanisms are not at all efficient in reactions involving oxirane. Differently, the reaction of \ce{^.CH} with cyclopentadiene leads to the formation of benzene and fulvenallene \citep{caster2021product}, and thus to ring expansion. This is somewhat expected because cyclopentadiene offers the possibility of forming conjugated species while expansion of oxirane can only lead to unstable four-membered rings.  This suggests that oxygen-bearing rings with more than three atoms might not be easily formed in the ISM, being a ring precursor more prone to open-chain products. The last comment concerns the most abundant product, \ce{HCO^. + C2H4}. Our investigation predicts that it further fragments into \ce{H^.} and CO. At the same time, the formation of the related product \ce{HOC^. + C2H4} is negligible, its rate coefficient being five order of magnitude smaller ($k \sim 10^{-15}$ cm$^3$ molecule$^{-1}$ s$^{-1}$). 

\section{Conclusions}

The ability to automatically and exhaustively explore reactive PESs is of crucial importance. The extreme conditions of the ISM favor reaction pathways that often deviate from terrestrial intuition, including barrierless associations, tunneling-driven processes, and complex multistep mechanisms (oxirane + \ce{^.CH} reaction providing a significant example). In addition to require a lot of human effort, manual exploration of complex PESs such as that of oxirane + \ce{^.CH} is both inefficient and prone to overlooking critical intermediates or transition states, hence leading to incomplete or biased reaction networks. By contrast, autonomous computational workflows capable of systematically and automatically mapping PESs can reveal a vast range of energetically accessible routes, thereby ensuring that astrochemical models can be based on exhaustive chemical networks. In this work, we demonstrated how this can be achieved with the algorithms available in the Chemoton software. Indeed, a fully automated workflow has been successfully applied to the exploration of the oxirane (c-\ce{C2H4O}) + \ce{^.CH} reaction for astrochemical purposes. The approach employed by Chemoton revealed a vast and highly connected reaction network (212 compounds, 818 reactions, more than 9300 elementary steps), thus demonstrating the capability of automated approaches to exhaustively map complex reactive landscapes. The methodology used in this work is able to provide a great starting point for the chemical exploration of entangled networks, such as those arising from the CH radical. 

Kinetic simulations indicate that the dominant pathway of the oxirane + \ce{^.CH} reaction is that leading to \ce{HCO^. + C2H4}, this accounts for about 85\% of the global rate. Thus, the main outcome of the title reaction is the opening of the oxirane ring. The formation of \ce{HOC^. + C2H4} is another possible destruction pathway for the oxirane ring but to a by far lesser extent. Minor channels are those leading to the formation of trans-propenal (6-8\%) and 2H-oxetene (4-5\%). Instead, production of methyl ketene contributes only to $\sim$0.5\%. While this reaction alone is not able to explain the trans-propenal/methyl ketene abundance ratio, when considered together with the \ce{CH3CHO + CH^.} process, it supports the idea that their relative abundance is governed by gas-phase kinetic effects.
The reaction between oxirane and \ce{^.CH} does not suggest any other relevant isomers of the \ce{C3H4O} family to be searched for in the ISM, but seems to propose that 2H-oxetene might be present in regions where oxirane is present. However, this species has never been studied by experimental rotational spectroscopy; therefore, a line catalog for its search is not available. 
Lastly, our results indicated the difficulty of increasing the ring size starting from oxirane, as the four-membered ring (2H-oxetene) is formed in small abundance and might undergo isomerization to acrolein under energetic conditions. This might contribute to explain why larger oxygen-bearing rings are not yet observed in the ISM.

\section*{Data Availability}
All data and software to reproduce the results of this study are available on Zenodo \cite{zenodoArchive}.

\begin{acknowledgments}
This work has been supported by MUR (PRIN Grant Numbers 202082CE3T, P2022ZFNBL and 20225228K5) and by the University of Bologna (RFO funds). 
The COST Action CA21101 ``COSY - Confined molecular systems: from a new generation of materials to the stars’’ is acknowledged. M.M. thanks the European Union -- Next Generation EU under the Italian National Recovery and Resilience Plan (PNRR M4C2, Investment 1.4 -- Call for tender n. 3138 dated 16/12/2021—CN00000013 National Centre for HPC, Big Data and Quantum Computing (HPC) -- CUP J33C22001170001).

\end{acknowledgments}

\clearpage

%\bibliography{references_abbrev}

\appendix

\renewcommand\thefigure{\thesection\arabic{figure}}
\setcounter{figure}{0}  
\section{Bimolecular Products}\label{app_bimolecular}

Figures showing the bimolecular products observed on the reactive \ce{C3H5O} PES and used in the MESS input, grouped on the base of their co-product. 

\begin{figure}[hb!]
    \centering
    \includegraphics[clip, trim=3cm 8.5cm 2.5cm 1cm, width=0.95\linewidth]{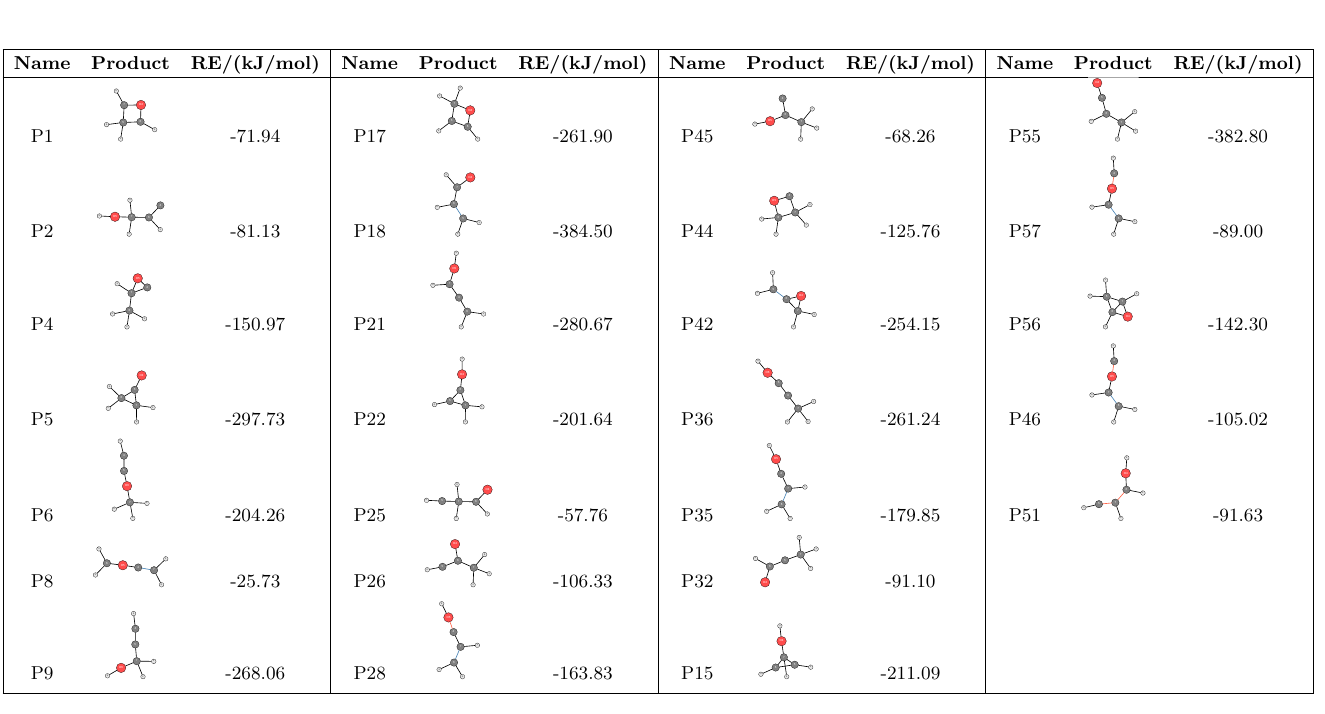}
    \caption{Bimolecular products having the H atom as co-fragment incorporated in the microkinetic modeling performed with MESS. Their ZPE-corrected relative energy (RE) with respect to the oxirane + \ce{^.CH} pair is reported in kJ/mol. ZPE correction is at the PBE-D3/def2-SVP level, electronic energy at the DLPNO-CCSD(T)-F12/cc-pVDZ-F12 level. Hydrogen atoms are shown in white, carbon atoms in grey, and oxygen atoms in red. Bonds that cannot be freely rotated, e.g., double or triple bonds, are colored.}
    \label{figA1}
\end{figure}

\clearpage

\begin{figure}
    \centering
    \includegraphics[clip, trim=2cm 10cm 2cm 1cm, width=0.95\linewidth]{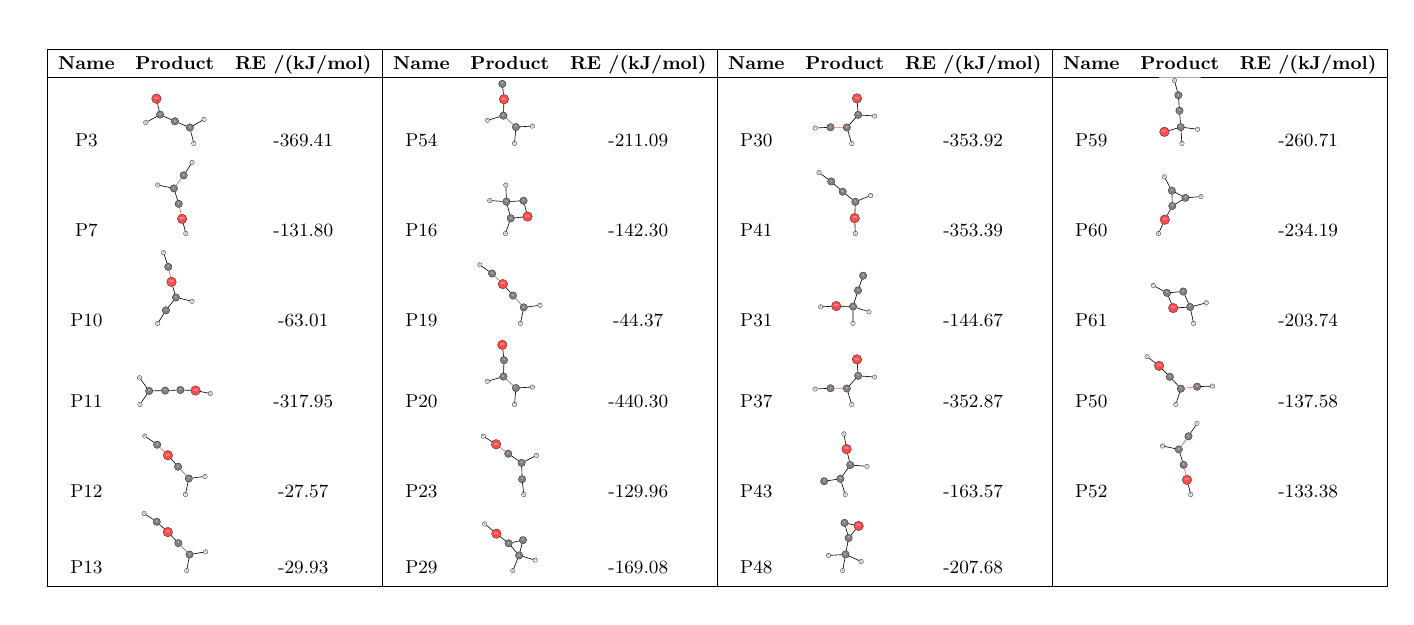}
    \caption{Bimolecular products having \ce{H2} as co-product incorporated in the microkinetic modeling performed with MESS. Their ZPE-corrected relative energy (RE) with respect to the oxirane + \ce{^.CH} pair is reported in kJ/mol. ZPE correction is at the PBE-D3/def2-SVP level, electronic energy at the DLPNO-CCSD(T)-F12/cc-pVDZ-F12 level. Hydrogen atoms are shown in white, carbon atoms in grey, and oxygen atoms in red. Bonds that cannot be freely rotated, e.g., double or triple bonds, are colored.}
    \label{figA2}
\end{figure}

\begin{figure}
    \centering
    \includegraphics[clip, trim=3cm 16.5cm 1cm 1cm, width=0.95\linewidth]{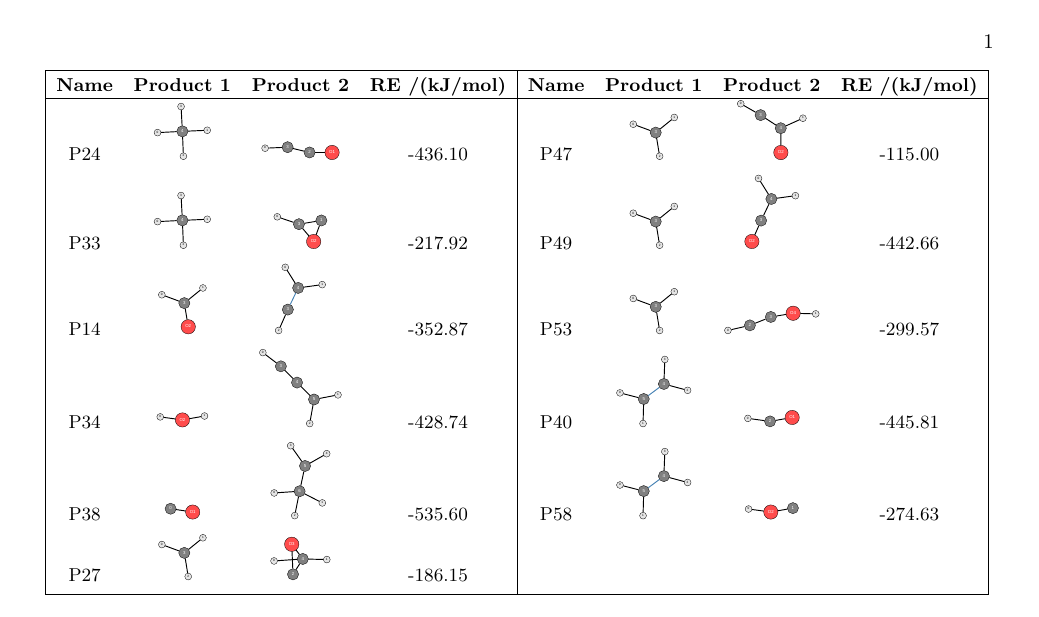}
    \caption{Bimolecular products with a co-fragment larger than H and \ce{H2} incorporated in the microkinetic modeling performed with MESS. Their ZPE-corrected relative energy (RE) with respect to the oxirane + \ce{^.CH} pair is reported in kJ/mol. ZPE correction is at the PBE-D3/def2-SVP level, electronic energy at the DLPNO-CCSD(T)-F12/cc-pVDZ-F12 level. Hydrogen atoms are shown in white, carbon atoms in grey, and oxygen atoms in red. Bonds that cannot be freely rotated, e.g., double or triple bonds, are colored.}
    \label{figA3}
\end{figure}

\clearpage

\section{Intermediates}\label{app_compounds}
\setcounter{figure}{0}  
In the following, the products and intermediates present in the Mater Equation simulations (i.e. those leading to bimolecular products lying below the reactants) is provided. For convenience of representation, the list of species is divided in four different figures, two for products and two for intermediates. For the \ce{C3H4O} isomers the co-product is \ce{H}, for the \ce{C3H3O} isomers is \ce{H2}. Noted is that, since only exothermic bimolecular products are considered, their number of compounds is lower than that reported in Tab.~\ref{tab:network_overview}. 

\begin{figure}[hb!]
    \centering
    \includegraphics[clip, trim=2.1cm 7cm 2.6cm 1cm, width=0.8\linewidth]{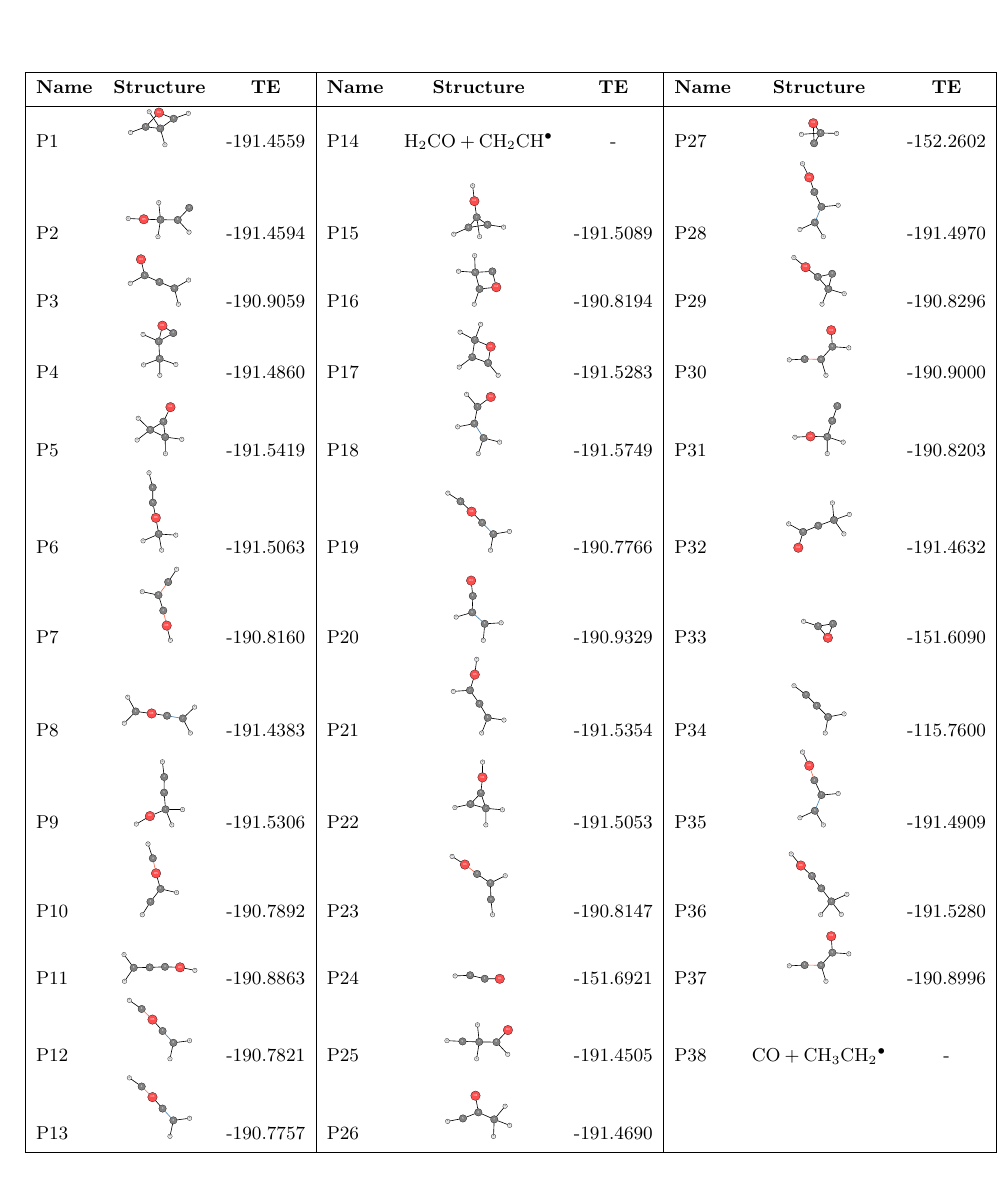}
    \caption{Products present in the MESS microkinetic modeling and their total energy (TE) in Hartree. The TE is obtained by combining the ZPE at the PBE-D3/def2-SVP level with the DLPNO-CCSD(T)-F12/cc-pVDZ-F12 electronic energy. Hydrogen atoms are shown in white, carbon atoms in grey, and oxygen atoms in red. Bonds that cannot be freely rotated, e.g., double or triple bonds, are colored.}
    \label{figC1}
\end{figure}

\begin{figure}[hb!]
    \centering
    \includegraphics[clip, trim=2.6cm 7cm 2.6cm 1cm, width=0.95\linewidth]{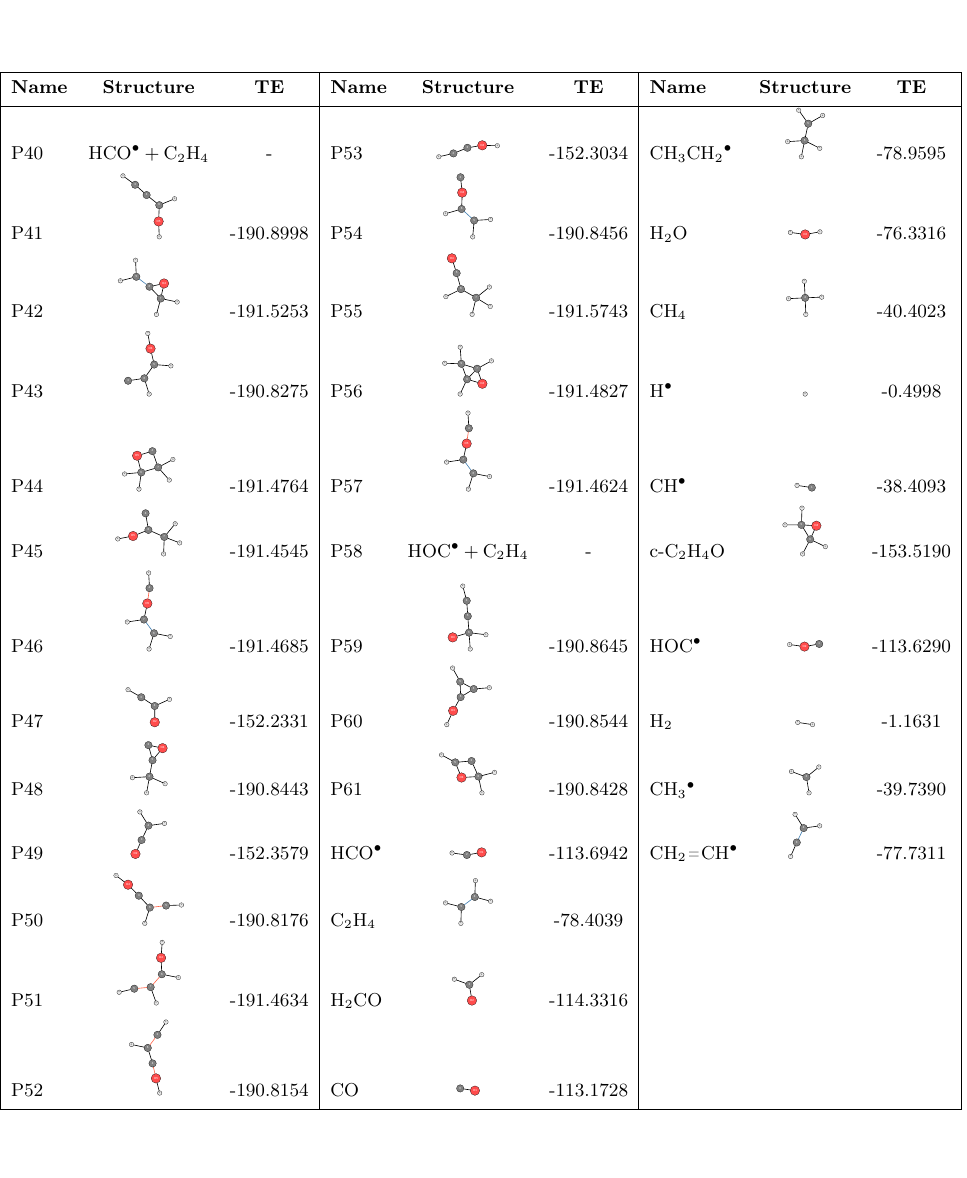}
    \caption{Products present in the MESS microkinetic modeling and their total energy (TE) in Hartree. The TE is obtained by combining the ZPE at the PBE-D3/def2-SVP level with the DLPNO-CCSD(T)-F12/cc-pVDZ-F12 electronic energy. Hydrogen atoms are shown in white, carbon atoms in grey, and oxygen atoms in red. Bonds that cannot be freely rotated, e.g., double or triple bonds, are colored.}
    \label{figC2}
\end{figure}

\begin{figure}[hb!]
    \centering
    \includegraphics[clip, trim=2.6cm 7cm 2.6cm 1cm, width=0.8\linewidth]{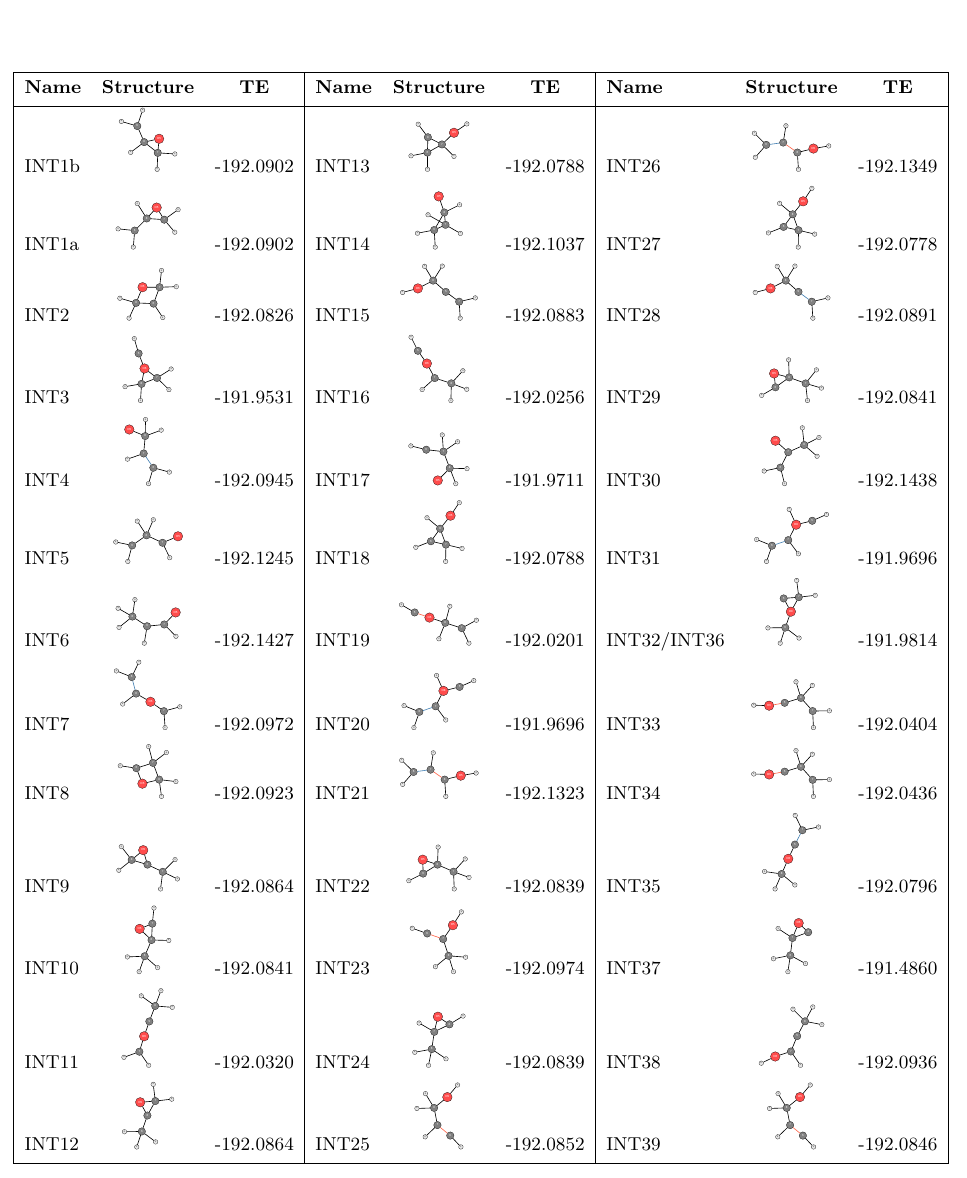}
    \caption{Intermediates used in the MESS microkinetic modeling and their total energy (TE) in Hartree. The TE is obtained by combining the ZPE at the PBE-D3/def2-SVP level with the DLPNO-CCSD(T)-F12/cc-pVDZ-F12 electronic energy. Hydrogen atoms are shown in white, carbon atoms in grey, and oxygen atoms in red. Bonds that cannot be freely rotated, e.g., double or triple bonds, are colored.}
    \label{figC3}
\end{figure}

\begin{figure}[hb!]
    \centering
    \includegraphics[clip, trim=4cm 7cm 2.6cm 1cm, width=0.8\linewidth]{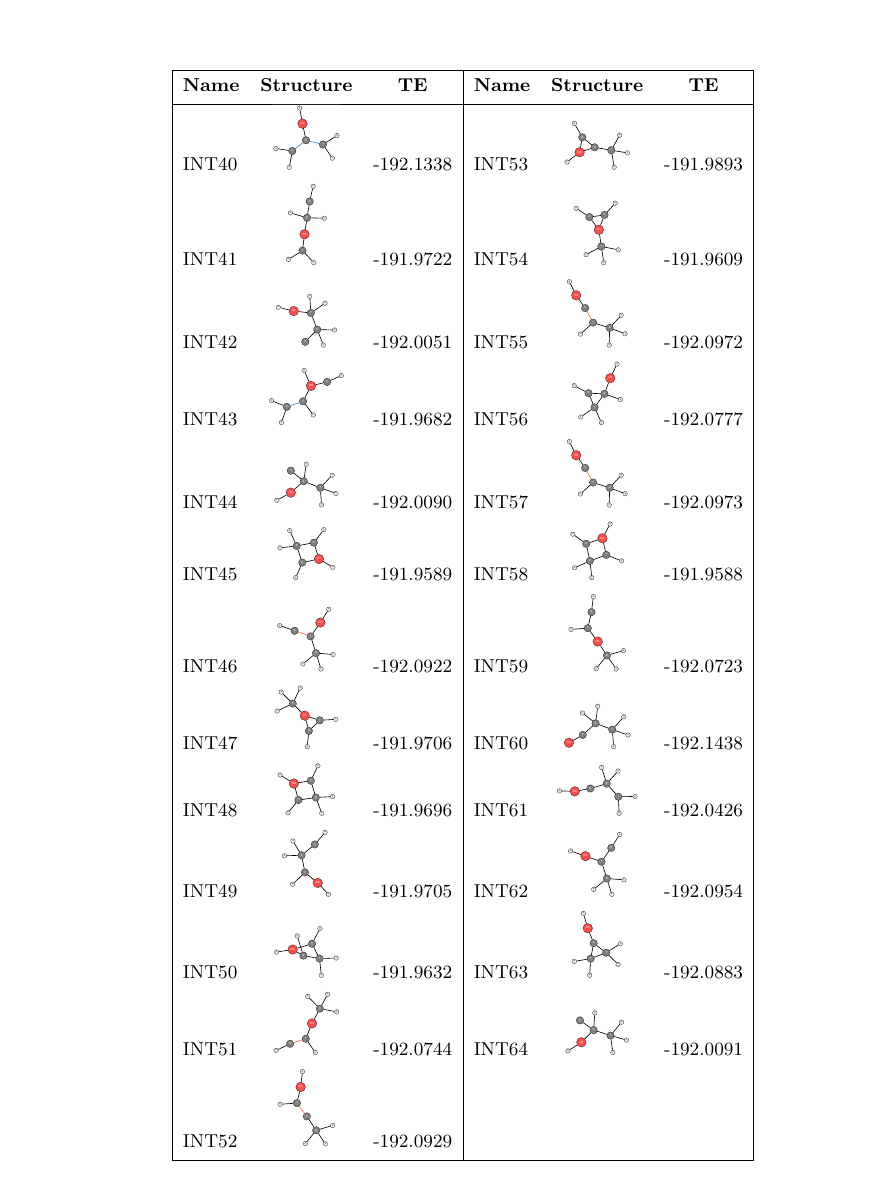}
    \caption{Intermediates used in the MESS microkinetic modeling and their total energy (TE) in Hartree. The TE is obtained by combining the ZPE at the PBE-D3/def2-SVP level with the DLPNO-CCSD(T)-F12/cc-pVDZ-F12 electronic energy. Hydrogen atoms are shown in white, carbon atoms in grey, and oxygen atoms in red. Bonds that cannot be freely rotated, e.g., double or triple bonds, are colored.}
    \label{figC4}
\end{figure}

\clearpage

\section{Entrance barrierless channels}
\label{app_entrancechannels}
\setcounter{figure}{0}  
The results of phase space theory (PST) applied to the barrierless entrance channels for the different approaches between oxirane and \ce{^.CH} are shown. 

\begin{figure}[h!]
    \centering
    \includegraphics[width=0.6\textwidth]{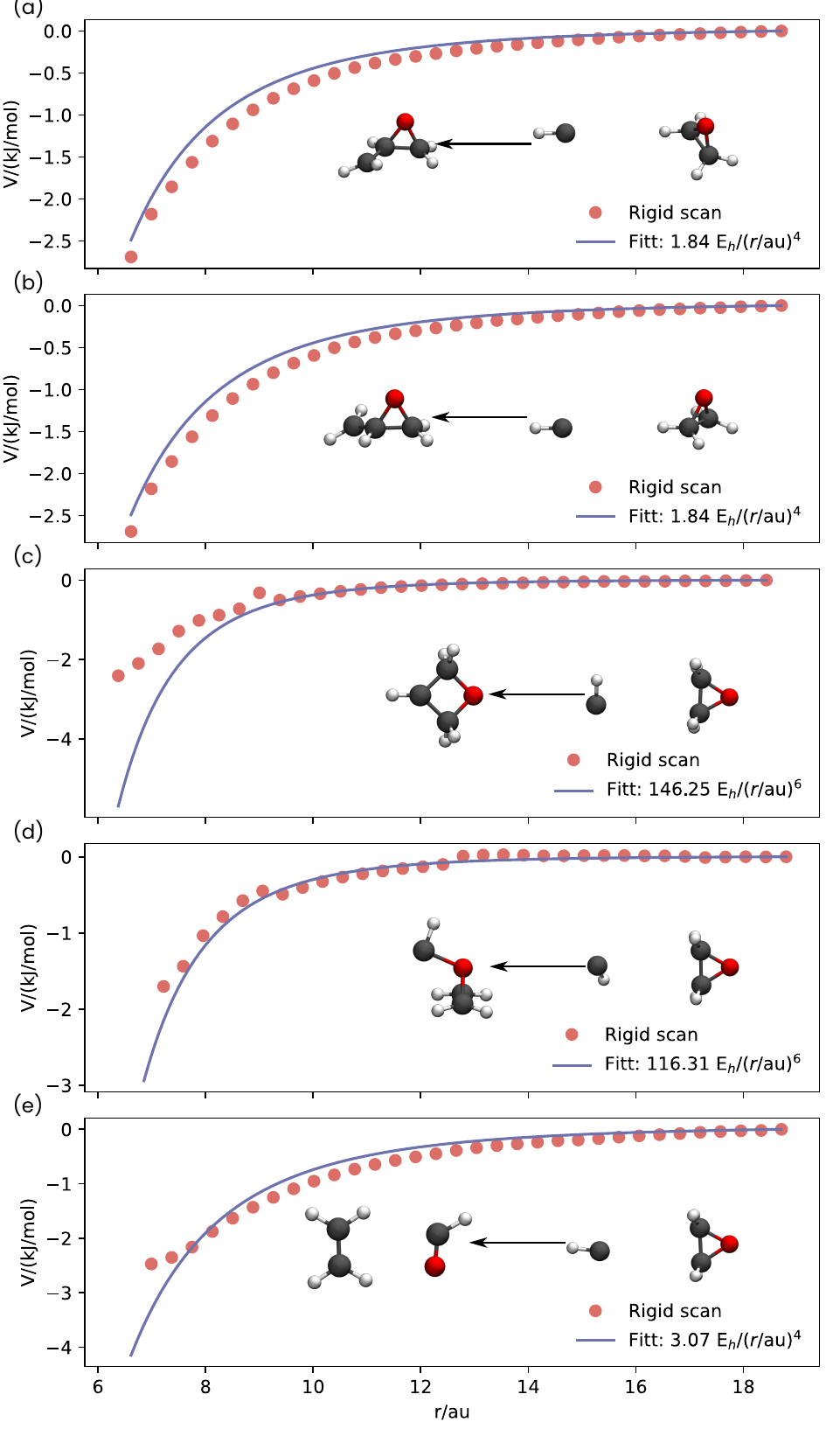}
    \caption{DLPNO-CCSD(T)-F12/cc-pVDZ-F12 potential energy ($V = E(r) - E(r\rightarrow\infty)$) along the barrierless entrance channels (a)-(e) and the corresponding fit of the PST parameters.}
    \label{fig:barrierless_channels}
\end{figure}

\clearpage

\section{Pressure-dependence Kinetics and Arrhenius-Kooji Fit}\label{pressure}
In this section, the results for the pressure dependence of the global rate constants and plots of the Arrhenius-Kooji fits are shown for the four main products of the oxirane + \ce{^.CH} reaction. The full range of temperature considered was used in the fitting procedure. 

\setcounter{figure}{0}  
\begin{figure}[h!]
    \centering
    \includegraphics[width=0.6\linewidth]{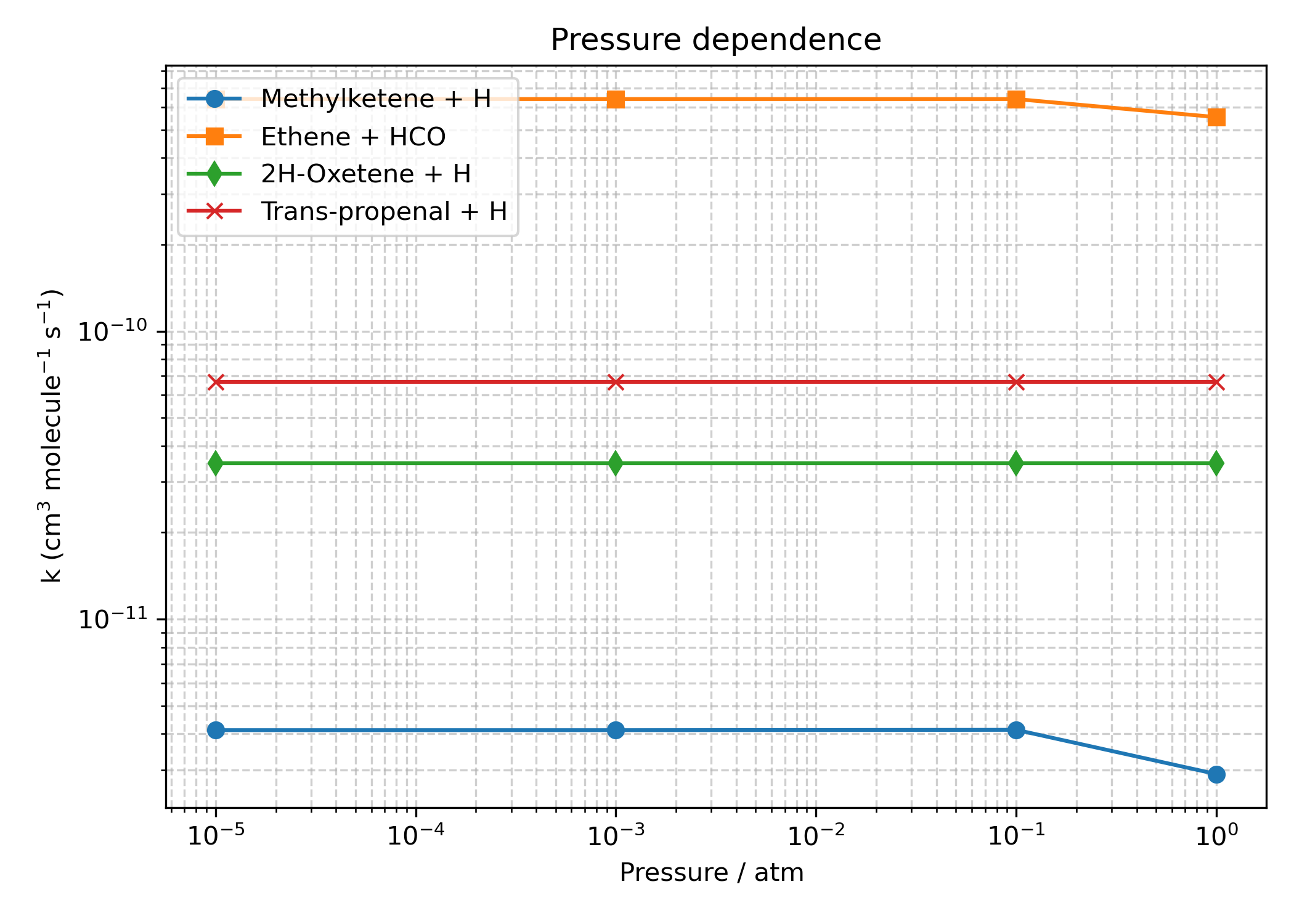}
    \caption{Pressure dependence of the reaction rates for the four main channels of the oxirane + \ce{^.CH} reaction.}
    \label{fig:pressure}
\end{figure}

\begin{figure}[h!]
    \centering
    \label{figurefit}
    \includegraphics[width=0.45\linewidth]{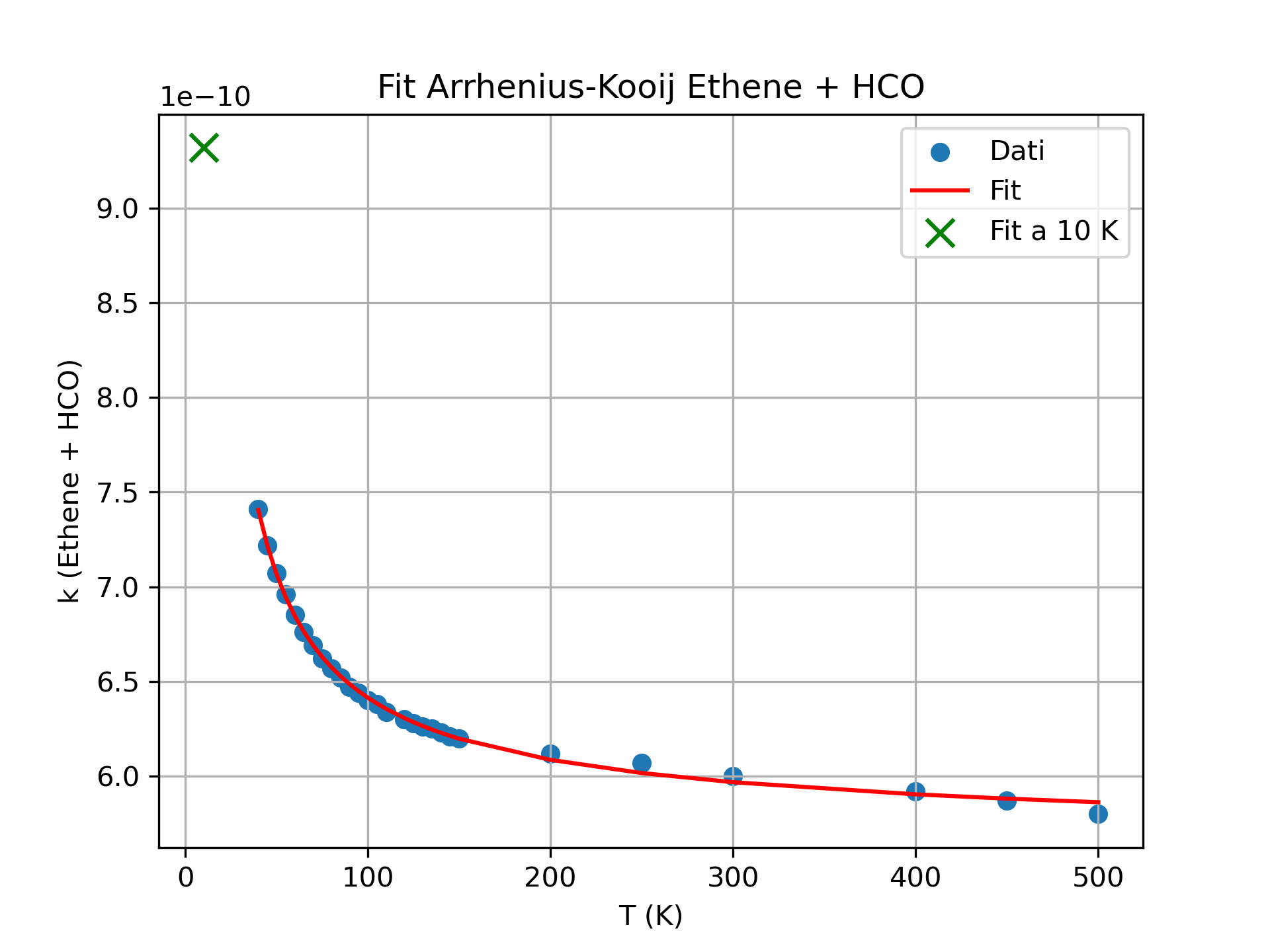}
    \includegraphics[width=0.45\linewidth]{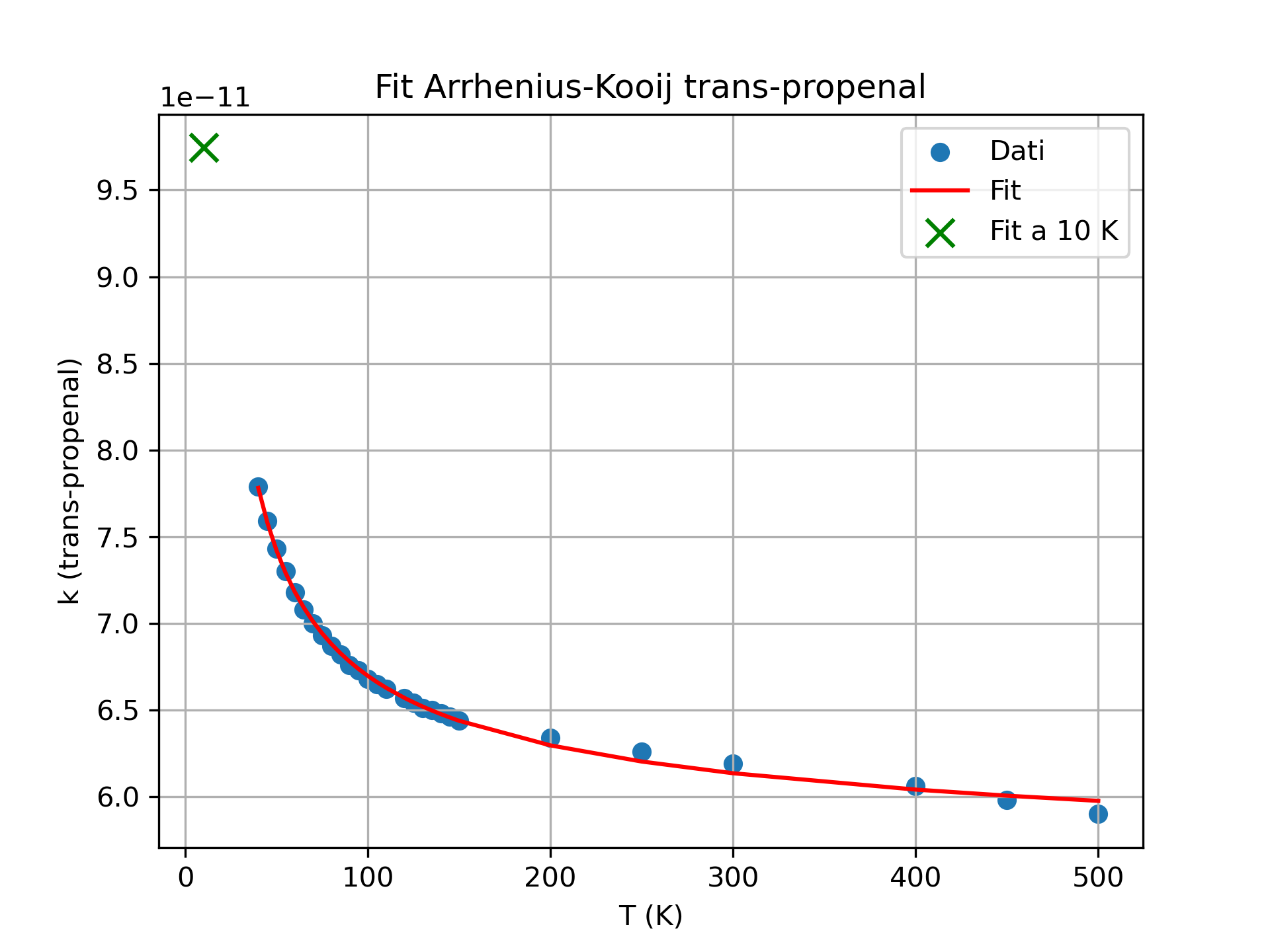}
    \includegraphics[width=0.45\linewidth]{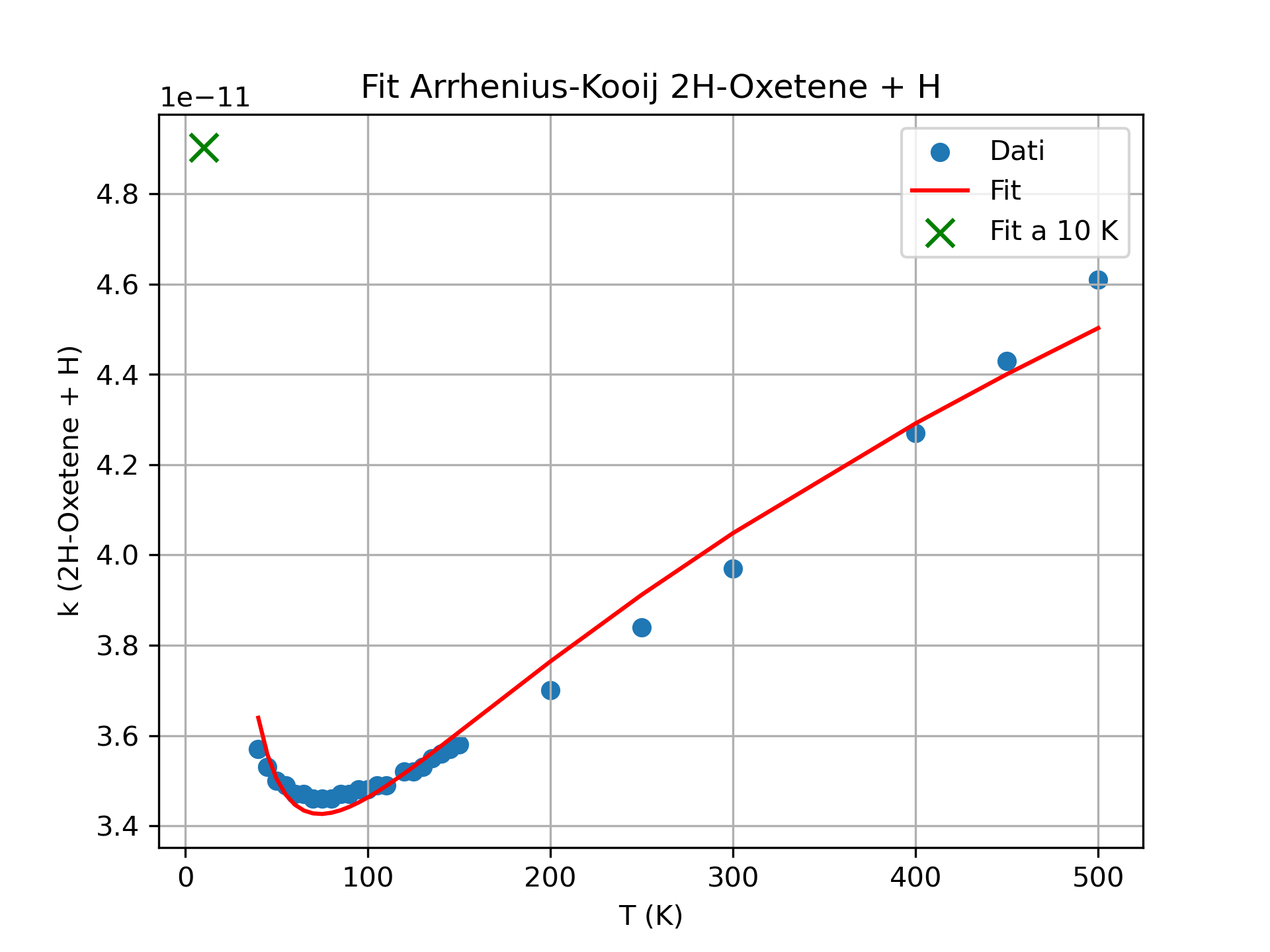}
    \includegraphics[width=0.45\linewidth]{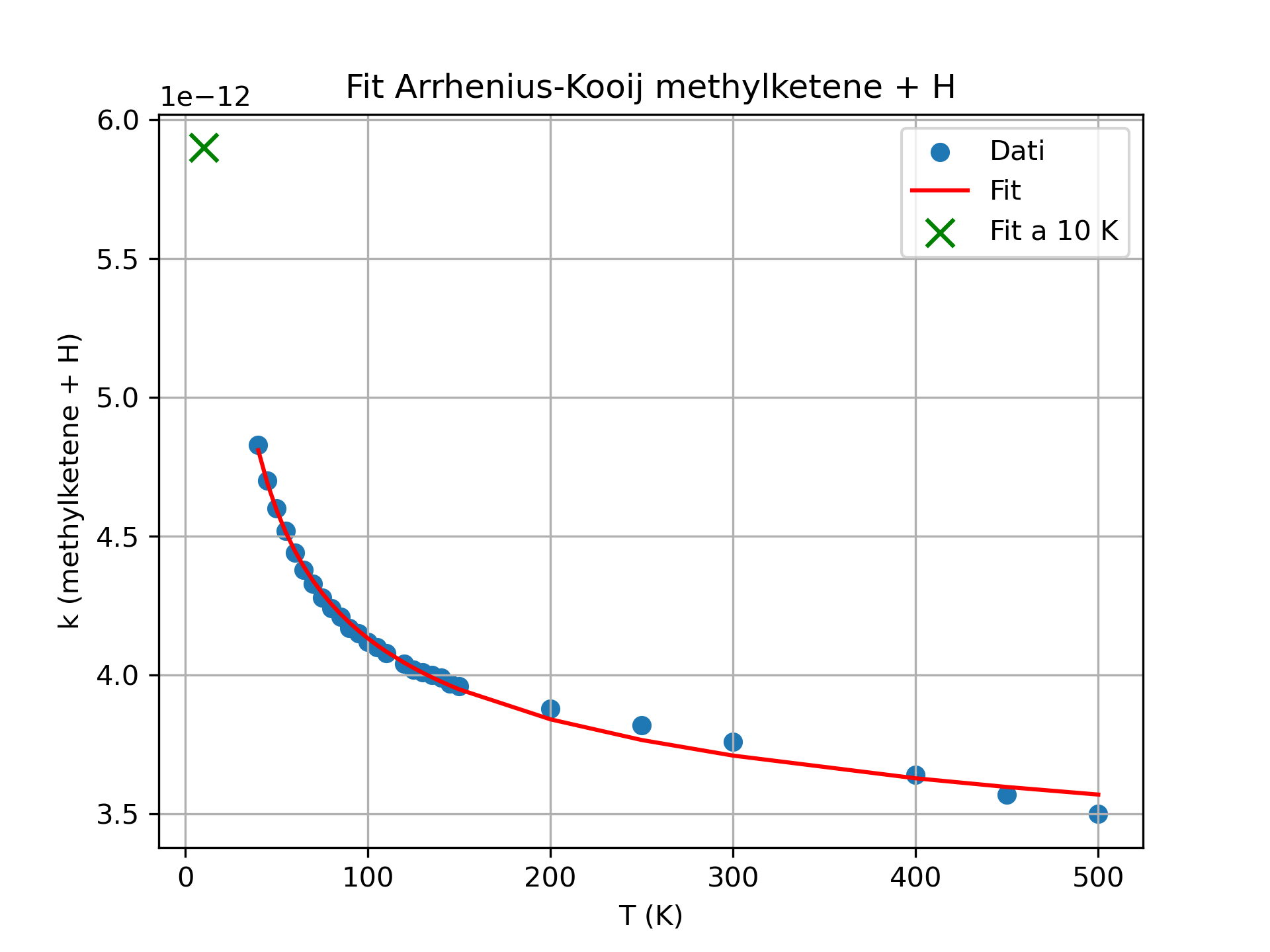}
    \caption{Plot of the rate coefficients (cm$^3$ molecule$^{-1}$ s$^{-1}$) for the four main reaction channels of the oxirane + \ce{^.CH} reaction and comparison with the Arrhenius-Kooji fit}
\end{figure}

\end{document}